\documentclass[a4paper,fleqn,11pt]{article}
\usepackage{amsmath}
\usepackage{amssymb}
\usepackage{hyperref}
\usepackage{pstricks}
\usepackage{graphicx}
\usepackage{cite}
\usepackage{multirow}
\usepackage{xspace}
\usepackage{ifthen}
\newcommand{\bea}{\begin{eqnarray}}
\newcommand{\eea}{\end{eqnarray}}
\newcommand{\nnb}{\nonumber}

\newcommand{\done}{{\rm d}}

\newcommand{\amegic}{A\scalebox{0.8}{MEGIC++}\xspace}

\newcommand{\sherpa}{S\scalebox{0.8}{HERPA}\xspace}
\newcommand{\vegas}{V\scalebox{0.8}{EGAS}\xspace}
\newcommand{\analysis}{A\scalebox{0.8}{NALYSIS}\xspace}
\newcommand{\disent}{D\scalebox{0.8}{ISENT}\xspace}

\makeatletter
\newif\if@preliminary
\@preliminaryfalse
\def\preliminary{\@preliminarytrue}
\def\preprintno#1{\def\@preprintno{#1}}
\def\address#1{\def\@address{#1}}
\def\email#1#2{\thanks{\tt #1@{}#2}}
\def\abstract#1{\def\@abstract{#1}}
\renewcommand\abstractname{ABSTRACT}
\newlength\preprintnoskip
\newlength\abstractwidth
\@titlepagetrue
\renewcommand\maketitle{\begin{titlepage}%
  \let\footnotesize\small
  \hfill\parbox{\preprintnoskip}{%
  \begin{flushright}\@preprintno\end{flushright}}\hspace*{1cm}
  \vskip 60\p@
  \begin{center}%
    {\Large\bf\boldmath \@title \par}\vskip 1cm%
    {\sc\@author \par}\vskip 3mm%
    {\@address \par}%
    \if@preliminary
      \vskip 2cm {\large\sf PRELIMINARY DRAFT \par \@date}%
    \fi
  \end{center}\par
  \@thanks
  \vfill
  \begin{center}%
    \parbox{\abstractwidth}{\centerline{\abstractname}%
    \vskip 3mm%
    \@abstract}
  \end{center}
  \end{titlepage}%
  \setcounter{footnote}{0}%
  \let\thanks\relax\let\maketitle\relax
  \gdef\@thanks{}\gdef\@author{}\gdef\@address{}%
  \gdef\@title{}\gdef\@abstract{}\gdef\@preprintno{}
}%
\def\@citex[#1]#2{\if@filesw\immediate\write\@auxout{\string\citation{#2}}\fi
  \def\@citea{}\@cite{\@for\@citeb:=#2\do
    {\@citea\def\@citea{,\penalty\@m}\@ifundefined
       {b@\@citeb}{{\bf ?}\@warning
       {Citation `\@citeb' on page \thepage \space undefined}}%
\hbox{\csname b@\@citeb\endcsname}}}{#1}}
\def\citerange{\@ifnextchar [{\@tempswatrue\@citexr}{\@tempswafalse\@citexr[]}}
\def\@citexr[#1]#2{\if@filesw\immediate\write\@auxout{\string\citation{#2}}\fi
  \def\@citea{}\@cite{\@for\@citeb:=#2\do
    {\@citea\def\@citea{--\penalty\@m}\@ifundefined
       {b@\@citeb}{{\bf ?}\@warning
       {Citation `\@citeb' on page \thepage \space undefined}}%
\hbox{\csname b@\@citeb\endcsname}}}{#1}}
\def\fmslash{\@ifnextchar[{\fmsl@sh}{\fmsl@sh[0mu]}}
\def\fmsl@sh[#1]#2{%
  \mathchoice
    {\@fmsl@sh\displaystyle{#1}{#2}}%
    {\@fmsl@sh\textstyle{#1}{#2}}%
    {\@fmsl@sh\scriptstyle{#1}{#2}}%
    {\@fmsl@sh\scriptscriptstyle{#1}{#2}}}
\def\@fmsl@sh#1#2#3{\m@th\ooalign{$\hfil#1\mkern#2/\hfil$\crcr$#1#3$}}
\makeatother

\begin{document}
\title{Automating dipole subtraction for QCD NLO calculations}
\preprintno{DCPT/07/88\\ IPPP/07/44}
\author{%
 T.~Gleisberg\email{tanju}{theory.phy.tu-dresden.de}$^a$,
 F.~Krauss\email{frank.krauss}{durham.ac.uk}$^b$
}
\address{
$^a$Institut f\"ur Theoretische Physik, D--01062 Dresden, Germany\\
$^b$Institute for Particle Physics Phenomenology, Durham University, Durham DH1 3LE, UK
\\[.5\baselineskip]
}
\abstract{
In this publication the construction of an automatic algorithm to subtract infrared
divergences in real QCD corrections through the Catani-Seymour dipole subtraction method 
\cite{Catani:1996vz} is reported.  The resulting computer code has been implemented
in the matrix element generator \amegic \cite{Krauss:2001iv}.  This will allow for the
automatic generation of dipole subtraction terms and their integrals over the one-parton 
emission phase space for any given process. If the virtual matrix element is provided as
well, this then directly leads to an NLO QCD parton level event generator. 
}
\maketitle
\pagenumbering{arabic}
\setcounter{page}{1}
\tableofcontents
\newpage
\section{Introduction}
\label{intro}

Perturbative calculations form one of the best understood methods to provide
predictions for the behavior of a Quantum Field Theory and to compare them
with experimental results.  Many of the methods applied in such calculations
have found their way into textbooks already decades ago, see e.g.\ 
\cite{Bjorken:1979dk}-\cite{Bohm:2001yx}.  Typically, the perturbation 
parameter is related to the coupling constant of the theory in question, 
which in most cases indeed is a small quantity.  This also implies that the 
corresponding fields may asymptotically appear as free fields and thus are 
the relevant objects of perturbation theory.  Obviously, this is not true 
for the strong interactions, i.e.\ QCD, where the fields, quarks and gluons, 
asymptotically are confined in bound states only.  This is due to the scaling 
behavior of the coupling constant of QCD, $\alpha_S$, which becomes small only 
for large momentum transfers, see for instance \cite{Ellis:1991qj,Brock:1993sz}.  
It is the confinement property that to some extent restricts the validity of 
perturbative calculations in QCD to the realm of processes characterized by large 
momentum transfers or by other large scales dominating the process, such that 
factorization theorems can be applied \cite{Brock:1993sz}-\cite{Collins:1989gx}.

Typically, for most of the relevant observables in particle phenomenology, the 
leading term of the perturbative expansion can be related to tree-level diagrams.  
In the past years, the calculation of these terms has been fully automated and a 
number of tools capable of dealing with up to eight to ten external particles 
without any significant user interference have emerged 
\cite{Krauss:2001iv,Kanaki:2000ey}-\cite{Maltoni:2002qb}.  
However, for many practical purposes, tree-level calculations are not sufficient.  
This is due to a number of reasons:  first of all, many measurements aim at the 
extraction of fundamental parameters.  However, in Quantum Field Theories, 
parameters are subject to corrections, which usually exhibit ultraviolet 
divergences.  These divergences are dealt with through the renormalization 
procedure, which can be done in a scheme- and scale-dependent way only, see 
e.g.\ \cite{Bardin:1999ak,Pascual:1984zb}.  Therefore, in order to extract 
parameters from the comparison of a (perturbative) calculation with experimental 
data, the calculation itself must contain the same kind of quantum corrections 
necessitating their renormalization.  Second, it should be stressed that in 
tree-level calculations, there are some choices to be made, concerning the 
scale at which inputs such as the coupling constant, quark masses or parton 
distribution functions are taken.  In principle, different scale choices are 
equivalent, and renormalization group theory guarantees that, when taking into 
account all orders, the effect of scale choices vanishes.  At leading order (LO), 
however, their impact may still be significant, such that tree-level calculations 
merely give the order of magnitude for corresponding cross sections etc.; a prime 
example for this is the production of a Higgs boson in gluon fusion processes,  
where only the next-to-next-to leading order correction significantly reduces the 
scale dependence and produces a stable result 
\cite{Harlander:2002wh,Anastasiou:2002yz}.  Thus, aiming at any more precise 
prediction, higher-order calculations are a crucial ingredient of phenomenological 
analyzes.    

But although indispensable, so far there is no fully automated tool available for 
QCD calculations at next-to leading order (NLO), i.e.\ at the one-loop level.  
This is because a true NLO calculation is certainly much more complex than a 
leading order (LO) one.  First of all, some of the essential ingredients, namely 
the loop or virtual contributions are not under full control yet.  In general, up 
to now calculations of these corrections to physical processes are limited to 
contributions containing five- and in some cases six-point functions, see for
example \cite{Campbell:2002tg}-\cite{Binoth:2007ca}.  But even to reach the level of 
known scalar master integrals is far from being trivial; the tensor reduction 
necessary for this step \cite{Passarino:1978jh} results in a proliferation of terms 
with non-trivial cancellations among them, which render the implementation in a 
computer code a major effort.  On the other hand, some of the loop corrections 
exhibit not only ultraviolet divergences to be renormalized, but also infrared 
divergences.  They also need to be regularized, but then they must be canceled 
against similar infrared divergences stemming from the real contributions.  This 
basically translates into canceling divergences in phase space volumes of different 
dimensionality.  The cancellation in fact is one of the most important consequences 
of the Kinoshita-Lee-Nauenberg or mass factorization theorems 
\cite{Bloch:1937pw}-\cite{Lee:1964is}.  However, in order to practically achieve 
the cancellation, the real infrared divergences also need to be regularized.  
Essentially, there are two ways of doing this.  

One method, also known as phase-space slicing \cite{Baer:1989jg}-\cite{Giele:1993dj}, 
bases on dividing the phase space of the additional real emission into an infrared-safe 
(hard) and a infrared-divergent (soft) region.  The division is usually performed by 
subjecting pairs of particles to an invariant mass criterion.  Then, the soft region 
is integrated analytically in $d$ dimensions.  Typically, in this step, the 
helicity-summed matrix element squared is approximated by its double-pole (or eikonal) 
limit.  The result of the analytical integration will contain single or double poles 
of the form $1/(d-4)$ or $1/(d-4)^2$, respectively.  They typically are accompanied 
with logarithms of the invariant mass criterion.  Such logarithms, but with opposite 
sign, also appear in the numerical evaluation of the full matrix element squared for 
real emission in the hard region of phase space, performed in 4 dimensions.  In 
principle, these two potentially large contributions (logarithms of a potentially 
small quantity) originate from the unphysical division of the phase space and should 
thus cancel.  Therefore, the key issue thus in phase space slicing is to adjust the 
parameters of the procedure such that the dependence on the slicing parameter is 
minimized.  So far, this adjustment has been done manually only and this is one of the 
reasons why other methods have become more popular with practitioners of NLO 
calculations.  

Such alternative methods of dealing with the real infrared divergences base on directly
subtracting them \cite{Catani:1996vz,Ellis:1980wv}-\cite{Somogyi:2006cz}\footnote{
        Subtraction methods for the NNLO case have been presented in 
        \cite{Kosower:2002su}-\cite{Daleo:2006xa}.}.  
At NLO level the subtraction in all methods is performed such that the additional 
particle is added to the leading order matrix element in a well-defined way through 
terms which, on one hand, exhibit the correct divergent behavior in the soft and 
collinear limit, and, on the other hand, can easily be integrated over the full 
$d$-dimensional phase space of the extra particle.  The idea is then that the so 
subtracted matrix element squared is finite and thus can safely be integrated numerically 
in 4 dimensions.  On the other hand,  the subtraction term is added to the virtual bit 
and integrated analytically in $d$ dimensions.  Again, it exhibits single or double 
poles of the form $1/(d-4)$ or $1/(d-4)^2$, respectively.  These poles again cancel 
the infrared poles of the virtual contributions.  The fact that there are universal 
subtraction terms, i.e.\ terms which will cancel the infrared divergences in a 
process-independent manner, is one of the main reasons why subtraction methods have 
become increasingly popular in past years and why they have been used for many of the 
state-of-the-art calculation of NLO corrections to physical processes, like for instance  
\cite{Campbell:2002tg}-\cite{Dittmaier:2007wz}.  

The universality of the subtraction terms also allow for an automated treatment of real 
infrared divergences.  It is the subject of this publication to report on a fully 
automated, process-independent implementation of one of the popular subtraction procedures, 
ready for use in realistic NLO calculations.  Therefore, the outline of this paper is as 
follows: In section 2, the anatomy of QCD NLO calculations will be formalized in a more 
mathematical language and the chosen subtraction method, the Catani-Seymour dipole 
subtraction \cite{Catani:1996vz} will briefly be reviewed in its original form for
massless particles.  Although its extension to massive particles \cite{Catani:2002hc} 
is straightforward from an algorithmic point of view, this paper concentrates on the
massless case only.  In section 3, the fully automated implementation of the corresponding 
massless dipole subtraction of arbitrary matrix elements into the matrix element generator 
\amegic \cite{Krauss:2001iv} will be presented in some detail.  Some simple tests of the 
implementation will be discussed in section 4, before some physical applications and 
the comparison with results from the literature will round of the presentation in section 5.

\section{Brief review of the Catani-Seymour formalism}

\subsection{NLO cross sections and the subtraction procedure}
\label{nloxs}
Cross sections at NLO precision are given by
\bea
\sigma=\sigma^{\rm LO}+\sigma^{\rm NLO}\;,
\eea
where the LO part $\sigma^{\rm LO}$ is obtained by integrating the exclusive cross section
in Born approximation over the available phase space of the $m$ final state particles and, 
eventually, over the Bjorken-$x$ of incident partons.  Ignoring this additional complication 
for the sake of a compact notation, The LO cross section is thus given by
\bea
\sigma^{\rm LO}=\int_{m}\done^{(4)}\sigma^{\rm B}\;,
\eea
where 
\bea
\done^{(4)}\sigma^{\rm B} = \done^{(4)}\Phi^{(m)}\left|{\cal M}_m\right|^2F_J^{(m)}\,.
\eea
Here, $\done^{(4)}\Phi^{(m)}$ denotes the phase space element of $m$ particles, taken in 
four dimensions, ${\cal M}_m$ is the matrix element for the process under consideration, 
and $F_J^{(m)}$ is a function of cuts defining the jets etc..  As already indicated, here
and in the following, the superscripts in the integral denote the dimensionality of the
integration.  In order to obtain a meaningful result to be compared with experimental data, 
typically isolation cuts are applied on the outgoing particles, which may also serve the 
purpose of keeping the integral finite.  A typical criterion for example is to identify 
outgoing partons with jets and thus apply jet definition cuts on the partons such that 
they are all well separated in phase space.  Anyway, the cuts will not be stated explicitly 
in the integral, but they are understood implicitly with the integration, including 
suitable generalisations in $d$ dimensions, where necessary.  Thus, the integration of 
the Born level cross section can directly be carried out in four space-time dimensions, 
as indicated in the equation.

In view of the dipole subtraction formulae, it is useful to introduce at this point bras and 
kets $\vphantom1_m\!\langle 1, \dots, m'|$ and $|1, \dots, m'\rangle_m$.  They denote states
of $m$ final state partons partons labeled by $1$ to $m'$ and are vectors in colour and 
helicity  space.  Introducing, in a similar fashion, vectors for the spins and colours, 
matrix elements thus can be written as
\bea
{\cal M}_m^{c_i,s_i} = 
\left(\vphantom1_m\!\langle c_1, \dots, c_m|\otimes
      \vphantom1_m\!\langle s_1, \dots, s_m|\right)|1, \dots, m\rangle_m\,.
\eea
Therefore, in this notation, the matrix element squared, summed over final state colours and 
spins reads
\bea
|{\cal M}_m|^2 = \vphantom1_m\!\langle 1, \dots, m|1, \dots, m\rangle_m\,.
\eea

The NLO part of the cross section consists of two contributions, each of which increases the 
order of $\alpha_S$.  First, there are emissions of an additional parton, i.e.\ real 
corrections, denoted by $\done^{(d)}\sigma^{\rm R}$.  Second, there are virtual (one-loop) 
corrections to the born matrix element, here denoted by $\done^{(d)}\sigma^{\rm V}$.  Thus,
\bea
\sigma^{\rm NLO} = \int\done^{(d)}\sigma^{\rm NLO} =
\int_{m+1}\done^{(d)}\sigma^{\rm R}+\int_{m}\done^{(d)}\sigma^{\rm V}\;.
\label{nloc}
\eea
The two integrals on the right-hand side of Eq. (\ref{nloc}) are separately infrared divergent 
in four dimensions, and are therefore taken in $d$ dimensions. For the real correction, the
divergences arise when the additional parton becomes soft or collinear w.r.t.\ some other parton,
leading to on-shell propagators in the matrix element.  For the virtual correction, the 
divergence comes with the integration over the unrestricted loop momentum, such that again a 
propagator goes on-shell.  
As already stated in the introduction, now the celebrated theorem of Kinoshita, Lee 
and Nauenberg \cite{Kinoshita:1962ur,Lee:1964is} comes to help and guarantees an exact 
cancellation of two divergent contributions, thus keeping their sum finite\footnote{
        In fact, this is only guaranteed for infrared-safe quantities.  More specifically, if 
        $F_J^{(n)}$ defines jets in terms of the momenta of an $m$-parton final state (taken 
        at Born level), infrared safety demands that $F_J^{(m+1)}\to F_J^{(m)}$ in cases where 
        the $m+1$- and $m$-parton configurations become kinematically degenerate.}.
Setting $d=4+2\epsilon$ in the following, the divergences will manifest themselves in double 
and single poles, i.e.\ as $1/\epsilon^2$ and $1/\epsilon$, respectively.  In principle, 
cancellation of the poles then solves the problem; in practice, however, the direct 
applicability of the equations above to real physical processes is limited since 
analytical integration over a multi-particle phase space in $d$ dimensions with cuts in 
many cases is beyond current abilities.

Therefore, a detour has to be taken.  The idea is to construct a subtraction term for the 
real emission contribution, which encodes all of its infrared divergences, but can 
analytically be integrated over in $d$ dimensions.  In this way the infrared pole structure 
of the real part with its $1/\epsilon$ and $1/\epsilon^2$ poles is exhibited and cancels 
the corresponding virtual contributions.  Subtracting this term from the real emission 
contribution and adding it to the virtual corrections then eliminates the infrared 
divergences in both parts.  The subtracted real matrix element squared then is finite and 
thus its full $(m+1)$-particle phase space can safely be integrated over in four dimensions.  
In this way, the subtraction term aims at an infrared regularisation of the two contributions 
at integrand level.
\bea
\sigma^{\rm NLO} &=&
\int_{m+1}\done^{(d)}\sigma^{\rm R}-\int_{m+1}\done^{(d)}\sigma^{\rm A}
+\int_{m+1}\done^{(d)}\sigma^{\rm A}+\int_{m}\done^{(d)}\sigma^{\rm V}\nnb\\ 
&\longrightarrow&
\int_{m+1}\left[\vphantom{\int}\done^{(4)}\sigma^{\rm R}-\done^{(4)}\sigma^{\rm A}\right]
+\int_{m+1}\done^{(d)}\sigma^{\rm A}+\int_{m}\done^{(d)}\sigma^{\rm V}\;.
\eea
The catch of the subtraction method now is that the subtraction terms can be obtained from 
the Born terms in a straightforward way and that only the phase space integral of the 
extra particle has to be taken in $d$ dimensions, while the phase space for the remaining 
$m$ particles can be taken in four dimensions.  This is similar to the way, the loop terms 
are evaluated.  There, only the loop integration is performed in $d$ dimensions, whereas 
the phase space of the outgoing particles is done in four dimensions.  Therefore, the 
final structure reads
\bea
\sigma^{\rm NLO}=
\int_{m+1}\left[\vphantom{\int}\done^{(4)}\sigma^{\rm R}-
                               \done^{(4)}\sigma^{\rm A}\right]
+\int_{m}\left[\int_{\rm loop}\done ^{(d)}\sigma^{\rm V}+
               \int_{1}\done^{(d)}\sigma^{\rm A}\right]_{\epsilon=0}\;.
\eea
Both integrands now are finite, allowing all integrations to be performed numerically.  In 
contrast to some other regularisation methods (like, e.g., phase space slicing) the subtraction 
method does not rely on any approximation and does not introduces any ambiguous and/or 
unphysical cut-off scales etc., as long as the integration of $\done^{(d)}\sigma^{\rm A}$ 
can exactly and analytically be performed.

In \cite{Catani:1996vz} a general expression for $\done^{(d)}\sigma^{\rm A}$ has been 
presented, called the dipole factorisation formula, allowing to write
\bea
\done^{(d)}\sigma^{\rm A} = 
\sum\limits_{\rm dipoles}\done^{(4)}\sigma^{\rm B}\otimes\done^{(d)}V_{\rm dipole}
\eea
such that, symbolically, 
\bea
\int_{m+1}\done^{(d)}\sigma^{\rm A} = 
\sum\limits_{\rm dipoles}\int_m\done^{(4)}\sigma^{\rm B}\otimes\int_1\done^{(d)}V_{\rm dipole} =
\int_m\left[\done^{(4)}\sigma^{\rm B}\otimes I\right]\;,
\eea
where
\bea
I = \sum\limits_{\rm dipoles}\int_1\done^{(d)}V_{\rm dipole}\;.
\eea
Here the sum of the dipole terms $V_{\rm dipole}$ contains all soft and collinear divergences 
of the real emission pattern.  This factorisation formula is suited for any process with 
massless partons, and fulfills all the requirements mentioned above.  An extension to massive 
partons has been presented in \cite{Catani:2002hc}.  

However, as already mentioned in the introduction, in this publication only the massless 
case will be considered.  In order to provide a self-contained description, all necessary 
analytic expressions will be listed in this publication.  

\subsection{Generalisation to hadronic initial states}

The cross sections discussed so far were given for point-like initial states.  For cross 
sections in hadron collisions, however, the differential cross sections above must be 
convoluted with parton distribution functions (PDFs):
\bea
\sigma(p,p^\prime) =
\sum_{a,b}\int_0^1 \done\eta f_a(\eta,\mu_F^2)
          \int_0^1 \done\eta^\prime f_b(\eta^\prime,\mu_F^2)
          \left[\sigma^{\rm LO}_{ab}(\eta p,\eta^\prime p^\prime)
               +\sigma^{\rm NLO}_{ab}(\eta p,\eta^\prime p^\prime,\mu_F^2)\right]\;.
\eea
Here the subscripts on the cross section denote the flavours of the incoming partons; for 
the total cross section a sum over them has to be performed.  For the NLO part, now the 
higher-order corrections residing in the PDFs must be taken care of.  This is done by 
supplementing the NLO part with a collinear subtraction term $d\sigma^C_{ab}$, such that
\bea
\sigma^{\rm NLO}_{ab}(p_a,p_b,\mu_F^2)=\int_{m+1}\done^{(d)}\sigma^R_{ab}(p_a,p_b)+
\int_{m}\done^{(d)}\sigma^V_{ab}(p_a,p_b)+
\int_{m}\done^{(d)}\sigma^C_{ab}(p_a,p_b,\mu_F^2)\;.
\eea
This new term contains collinear singularities, incorporated in $1/\epsilon$-terms and reads
\bea\label{cct}
\done^{(d)}\sigma^C_{ab}(p_a,p_b,\mu_F^2) &=&
-\frac{\alpha_S}{2\pi}\frac1{\Gamma(1-\epsilon)}\sum_{c,d}
   \int_0^1\done z\int_0^1\done\bar{z}\left\{\vphantom{\frac{|^|}{|^|}}
   \done^{(4)}\sigma^B_{cd}(zp_a,\bar{z}p_b)\right.\nnb\\
&&\hspace*{8mm}
\cdot\left[\delta_{bd}\delta(1-\bar{z})\left(
   -\frac1\epsilon\left(\frac{4\pi\mu^2}{\mu_F^2}\right)^\epsilon 
   P_{ac}(z)+K^{\rm F.S.}_{ac}(z)\right)\right.\nnb\\
&&\hspace*{8mm}\left.\vphantom{\frac{|^|}{|^|}}\left.
+\delta_{ac}\delta(1-z)\left(
   -\frac1\epsilon\left(\frac{4\pi\mu^2}{\mu_F^2}\right)^\epsilon 
   P_{bd}(\bar{z})+K^{\rm F.S.}_{bd}(\bar{z})\right)\right]\right\}\;.
\eea
The collinear subtraction term is factorisation-scale and scheme dependent.  
This scheme dependence resides in the terms $K^{\rm F.S.}$, which, for the common 
$\overline{MS}$-scheme vanish, i.e.\ in this scheme all terms $K^{\rm F.S.}=0$.  
However, this scheme dependence cancels similar terms in the PDFs such that, 
taken together, the full hadronic cross section again is scheme-independent.

In the case of incoming hadrons, the subtraction method is applied to  
$\sigma^{\rm NLO}(p_a,p_b,\mu_F^2)$ as described before, with the only difference 
that in this case the singularities of $\done\sigma^V_{ab}$ only cancel in the sum
\bea
\int_{m}\left[\int_{\rm loop}\done^{(d)}\sigma^{\rm V}_{ab}+
              \int_{1}\done^{(d)}\sigma^{\rm A}_{ab}+
              \done^{(d)}\sigma_{ab}^{\rm C}\right]_{\epsilon=0}\;.
\eea

\subsection{Observable-independent formulation of the subtraction method}
\label{nloobservable}
Up to now, the $\done\sigma$ denoted cross sections in a broad sense.  To be a bit more 
specific consider the following expression for a cross section at Born-level and the 
corresponding next-to leading order expression:
\bea
\sigma^{\rm LO}
&=&
\int\done\Phi^{(m)}(p_1,...,p_m)\; 
    \left|M^{(m)}(p_1,...,p_m)\right|^2\; F^{(m)}(p_1,...,p_m)\nnb\\
\sigma^{\rm NLO}
&=& 
\int\done\Phi^{(m+1)}(p_1,...,p_{m+1})\; 
    \left|M^{(m+1)}(p_1,...,p_{m+1})\right|^2\;F^{(m+1)}(p_1,...,p_{m+1})\nnb\\
&&+
\int\done\Phi^{(m)}(p_1,...,p_m)\;\left|V^{(m)}(p_1,...,p_m)\right|^2\; F^{(m)}(p_1,...,p_m)\;,
\label{genobs}
\eea
where $\done\Phi^{(n)}$ represents an $n$-particle phase space element, and $M^{(m)}$, 
$M^{(m+1)}$ and $V^{(m)}$ are the LO matrix element, the NLO real matrix element and the 
NLO virtual correction matrix element, respectively.  $F^{(n)}$ is a function that defines 
a cross section or an observable in terms of the $n$-parton momentum configuration.  In 
general, the function $F$ may contain $\theta$-functions (to define cuts and corresponding 
total cross sections), $\delta$-functions (defining differential cross sections), kinematic 
factors or any combination of these.

However arbitrary this sounds, there is a formal requirement on this function $F$, namely
that in the soft and collinear limits, i.e.\ for cases where one parton becomes collinear
w.r.t.\ another one or where one parton becomes soft, the function $F^{(m+1)}$ reduces
to $F^{(m)}$:
\bea
F^{(m+1)}(p_1,...,p_i=\lambda q,...,p_{m+1}) 
&\to& 
F^{(m)}(p_1,...,p_{m+1})\;\; {\rm for}\;\lambda\to 0\nnb\\
F^{(m+1)}(p_1,...,p_i,...,p_j,...,p_{m+1})
&\to& 
F^{(m)}(p_1,...,p,...,p_{m+1})\;\; {\rm for}\; p_i\to zp,\;p_j\to (z-1)p\nnb\\
F^{(m)}(p_1,...,p_m)&\to& 0\;\; {\rm for}\;p_i\!\cdot\! p_j\to0\;.
\eea
The first two conditions define infrared-safe observables - to phrase it intuitively this
means that such infrared-safe quantities must not be altered by additional soft or collinear 
activity.  The last condition above is required to properly define the Born cross section.

Applying the subtraction method to the NLO-part of Eq.(\ref{genobs}) results in
\bea
\label{observablereq}
\sigma^{\rm NLO}&=&\;\;\;\;
\int\done\Phi^{(m+1)}\; 
    \left[\left|M^{(m+1)}(p_1,...,p_{m+1})\right|^2\; F^{(m+1)}(p_1,...,p_{m+1})
                                                 \vphantom{\sum_{k\neq i\neq j}}\right.\nnb\\
&& \hphantom{aaaaaaaaa}
\left. -\sum_{k\neq i\neq j}{\cal D}_{ij,k}(p_1,...,p_{m+1})\; 
                            F^{(m)}(p_1,..,\tilde{p}_{ij},\tilde{p}_k,..,p_{m+1})\right]\nnb\\
&&+\int\done\Phi^{(m)}\; \left[\left|V^{(m)}(p_1,...,p_m)\right|^2\;
                               +\left(\int_1\done[\tilde{p}] 
                                            {\cal D}_{ij,k}(p_1,...,p_{m+1})\right)\;
                                                 \vphantom{\sum_{k\neq i\neq j}}
                          \right] 
                         F^{(m)}(p_1,...,p_m)\;,\nnb\\
\eea
where $\done[\tilde{p}]$ is the phase space element for the 1-parton phase space.

In order to have an identity between the subtracted terms and the added term, both the 
$(m+1)$-parton contribution and the $m$-parton contribution have to be subjected to the 
same function $F$.  To be able to perform the integration over the one-parton phase 
space independent of the observable this function therefore must be $F^{(m)}$. In the 
case of the $(m+1)$-parton contribution $F^{(m)}$ is applied to the $m$-parton 
configuration, generated by corresponding mapping given in the prescription of the 
dipole function. 

\subsection{The dipole subtraction functions}
\label{subtraction}
The universality of the soft and collinear limits of QCD matrix elements are the basis 
for the construction of the dipole subtraction terms.  In both limits any matrix 
element squared for $m+1$-partons factorizes into an $m$-parton matrix element times 
a (singular) factor.

To be specific, consider first the soft limit of the matrix element, given by the momentum 
$p_j$ of parton $j$ becoming soft, i.e.\ $p_j^\mu=\lambda q^\mu$ with $\lambda\to0$
\cite{}.  Then, employing
\bea
\frac{p_ip_k}{(p_iq)(p_kq)} = 
\frac{p_ip_k}{(p_iq)[(p_i+p_k)q]} + \frac{p_ip_k}{[(p_i+p_k)q](p_kq)}\,,
\eea
the soft limit reads
\bea
\lefteqn{
{\vphantom1}_{m+1}\!\langle 1, \dots, j, \dots,m+1|
                            1, \dots, j, \dots,m+1\rangle_{m+1}}\nnb\\
&\longrightarrow&
-\frac1{\lambda^2}8\pi\mu^{2\epsilon}\alpha_S
     \sum_{i,k\neq i} 
     {\vphantom{\left\langle\frac{|}{|}\right\rangle}}_{m}\!\!
                \left\langle 1, \dots, i, \dots, m+1\left|
                                    \frac{p_kp_i{\bf T}_k\cdot{\bf T}_i}{(p_iq)[(p_i+p_k)q]}
                         \right|1, \dots, k, \dots, ,m+1\right\rangle_{m}\;.
\label{soft}
\eea
In a similar way the limit where two partons $i$ and $j$ become collinear is defined
through $p_j\to (1-z)/z\; p_i$.  In this limit the $(m+1)$ parton matrix element can be 
rewritten as
\bea
\lefteqn{{\vphantom1}_{m+1}\!\langle 1,...,m+1||1,...,m+1\rangle_{m+1}}\nnb\\
&\longrightarrow& \frac{1}{p_ip_j}\;4\pi\mu^{2\epsilon}\alpha_S\;
{\vphantom{\left\langle\frac{a}{a}\right\rangle}}_{m}\!\!
\left\langle 1, \dots ,m+1\left|\hat P_{(ij),i}(z,k_\perp)
                    \right|1, \dots ,m+1\right\rangle_{m}\;,
\eea
where, again, the $\hat P_{(ij),i}(z,k_\perp)$ are the well-known Altarelli-Parisi splitting 
functions.

Then, the actual dipole function generating the limit, where one of the partons $i,j$ of a 
$m+1$-parton configuration becomes soft or both partons become collinear to each other, 
symbolically has the following structure:
\bea\label{Eq:OneDipoleSubtractionTerm}
{\cal D}_{ij,k}=
{\vphantom1}_{m}\!\langle 1,...,\tilde{ij},...,\tilde{k},...,m|
                  |1,...,\tilde{ij},...,\tilde{k},...,m\rangle_{m}\;\otimes{\bf V}_{ij,k}\;,
\eea
with the non-singular $m$-parton matrix element 
${\vphantom1}_{m}\!\langle ...||...\rangle_{m}$ and the operator ${\bf V}_{ij,k}$, 
describing the splitting of the parton $(ij)$.  Here, and in the following, the 
splitting kernels ${\bf V}_{ij,k}$ are matrices in the helicity space of the emitter.  
The dipole function also involves a third parton as 'spectator'.  This parton in fact 
is identical with the colour partner $k$ in the soft limit, Eq.\ (\ref{soft}).  The 
form of the subtraction means that kinematically, $3\to 2$ mappings are considered
\bea
p_i,p_j,p_k\to \tilde p_{ij}, \tilde p_k\;,
\eea 
such that all involved partons are allowed to remain on their mass shells. 

\begin{figure}
  \begin{center}
    \includegraphics[width=150mm]{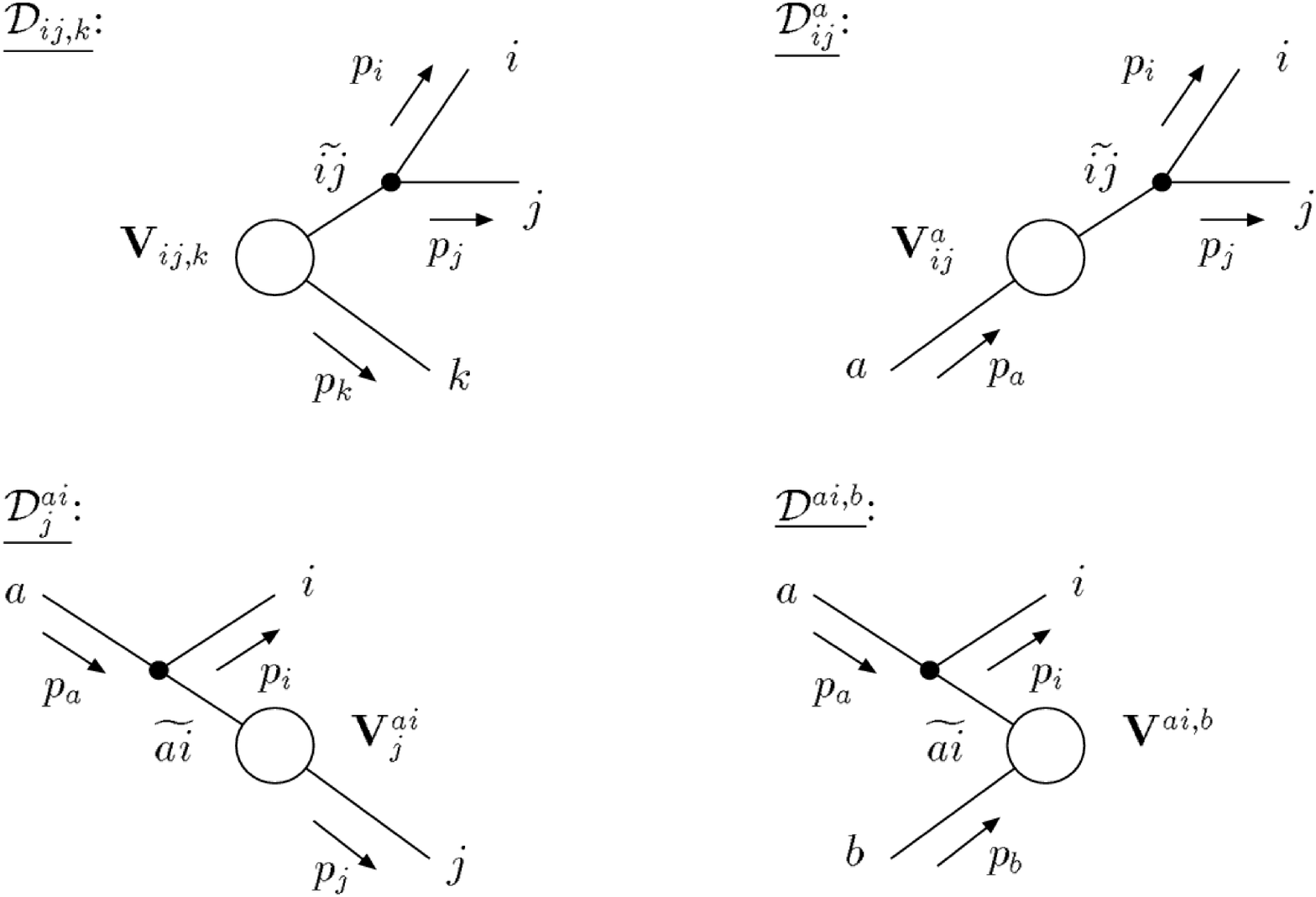}
    \caption{\label{dipoles}Classification of dipole functions.}
  \end{center}
\end{figure}

In general the splitting parton (called 'emitter') and the spectator can be both, 
initial and final state particles.  This discriminates four different types of dipole 
functions, displayed in Fig.\ \ref{dipoles}.  

The full subtraction term for any matrix element with $(m+1)$ partons in the final 
state is given by the sum of all possible dipole functions.  For the most general case 
with two partons in the initial state, therefore
\bea
\done\sigma^{A}=\left[
\sum_{k\neq i\neq j}{\cal D}_{ij,k}
+\left\{\sum_{i\neq j}{{\cal D}_{ij}}^a
       +\sum_{k\neq i}{{\cal D}_{k}}^{ai}
       +\sum_{i}{\cal D}^{ai,b}+(a\leftrightarrow b)\right\}\right]d\Phi^{(m+1)}\;.
\eea

In the following the explicit expressions for the dipole functions will be listed.  The 
corresponding one-parton phase space integrated subtraction terms are discussed in Sec.\ 
\ref{finitet}.

\subsubsection{Final state emitters with final state spectators}
\label{sff}
The dipole contribution ${\cal D}_{ij,k}$ for the singular limit $p_i\!\cdot\!p_j\to0$, 
where all three involved partons are in the final state, is given by
\bea
\label{ffdipole}
\lefteqn{{\cal D}_{ij,k}(p_1,\ldots,p_{m+1})}\nnb\\
&=&
-\frac1{2p_i\!\cdot\!p_j}
{\cdot\;}
     {\vphantom{\left\langle\frac{|_|}{|_|}\right\rangle}}_{m}\!\!
               \left\langle 1,\ldots,\tilde{ij},\ldots,\tilde{k},\ldots,m+1\left|
               \frac{{\bf T}_k\cdot{\bf T}_{ij}}{{\bf T}^2_{ij}}{\bf V}_{ij,k}
               \right|1,\ldots,\tilde{ij},\ldots,\tilde{k},\ldots,m+1\right\rangle_m\;.\nnb\\
\eea
It is obtained from an $(m+1)$-parton matrix element by replacing the partons $i$ and $j$ 
with a single parton $\tilde{ij}$, the emitter, and the parton $k$ is replaced by $\tilde{k}$,
the spectator.  The flavours of emitter and spectator are assigned as follows: The spectator
$\tilde{k}$ remains unchanged, and the emitter $\tilde{ij}$ is defined by the splitting
process $\tilde{ij}\to i+j$.  The product of colour charges in the numerator of Eq.\ 
(\ref{ffdipole}) introduces an extra colour correlation in the $m$-parton matrix element.  

The kinematics of the splitting are described by the following variables
\bea
y_{ij,k}=\frac{p_ip_j}{p_ip_j+p_jp_k+p_kp_i}\;,\;\;\;
\tilde{z_i}=\frac{p_ip_k}{p_jp_k+p_ip_k}=1-\tilde{z_j}\;.
\eea
and to obtain the momenta $\tilde{ij}$ and $\tilde{k}$ in the $m$-parton configuration
the following map is being used:
\bea
\tilde{p}_k^\mu=\frac1{1-y_{ij,k}}p_k^\mu\;,\;\;\;
\tilde{p}_{ij}^\mu=p_i^\mu+p_j^\mu-\frac{y_{ij,k}}{1-y_{ij,k}}p_k^\mu\;.
\eea
Obviously, four-momentum conservation is exactly fulfilled, i.e.\
\bea
p_i^\mu+p_j^\mu+p_k^\mu=\tilde{p}_{ij}^\mu+\tilde{p}_k^\mu
\eea
and all partons remain on their mass shell,
\bea
p_i^2 = p_j^2 = p_k^2=\tilde{p}_{ij}^2 = \tilde{p}_k^2 = 0\,.
\eea

The splitting matrices, which are related to the $d$-dimensional Altarelli-Parisi splitting
functions, depend on the spin indices of the emitter parton.  For the case of a quark 
splitting (using helicity indices $s$ and $s^\prime$) the kernel is a matrix in helicity
space, whereas for gluon splittings (to a quark-anti-quark pair or to gluons), the splitting 
matrices are given by Lorentz tensors.  This yields
\bea
\langle s|V_{q_ig_j,k}(\tilde{z_i};y_{ij,k})|s^\prime\rangle
&=&
8\pi\mu^{2\epsilon}\alpha_SC_F\left[
\frac2{1-\tilde{z_i}(1-y_{ij,k})}-(1+\tilde{z_i})-\epsilon(1-\tilde{z_i})
\right]\delta_{ss^\prime}\;,\nnb\\
\langle \mu|V_{q_i\bar{q}_j,k}(\tilde{z_i};y_{ij,k})|\nu\rangle
&=&
8\pi\mu^{2\epsilon}\alpha_ST_R\left[
-g^{\mu\nu}-\frac2{p_ip_j}(\tilde{z_i}p_i-\tilde{z_j}p_j)^\mu
                          (\tilde{z_i}p_i-\tilde{z_j}p_j)^\nu\right]\;,\nnb\\
\langle \mu|V_{g_ig_j,k}(\tilde{z_i};y_{ij,k})|\nu\rangle
&=&
16\pi\mu^{2\epsilon}\alpha_SC_A\left[
-g^{\mu\nu}\left(
\frac1{1-\tilde{z_i}(1-y_{ij,k})}+\frac1{1-\tilde{z_j}(1-y_{ij,k})}-2\right)\right.\nnb\\
&&\hspace*{22mm}\left.
+(1-\epsilon)\frac2{p_ip_j}(\tilde{z_i}p_i-\tilde{z_j}p_j)^\mu
                           (\tilde{z_i}p_i-\tilde{z_j}p_j)^\nu\right]\;,
\eea
respectively.  The dipole terms given in this section are sufficient for the subtraction 
procedure in the case of non-hadronic initial states such as $e^-e^+$-annihilation.

\subsubsection{Final state emitters with initial state spectators}
For the case of an emitting final state parton, the presence of an initial state 
spectator results in additional contributions to the singular limit $p_i\!\cdot\!p_j\to0$ 
of the full $m+1$-parton matrix element.  The corresponding dipole terms in this case are 
given by
\bea
\label{fidipole}
\lefteqn{{\cal D}_{ij}^a(p_1,\ldots,p_{m+1};p_a,..)}\nnb\\
&=&
-\frac1{2p_i\!\cdot\!p_j}\frac1{x_{ij,a}}
{\cdot\;}
     {\vphantom{\left\langle\frac{|_|}{|_|}\right\rangle}}_{m,a}\!\!
\left\langle 1,\ldots,\tilde{ij},\ldots,m+1;\tilde{a},..\left|
             \frac{{\bf T}_a\cdot{\bf T}_{ij}}{{\bf T}^2_{ij}}{\bf V}_{ij}^a
             \right|1,\ldots,\tilde{ij},\ldots,m+1;\tilde{a},..\right\rangle_{m,a}\,.\nnb\\
\eea
The kinematic variables now read
\bea
x_{ij,a}=1-\frac{p_ip_j}{(p_i+p_j)p_a}\;,\;\;\;
\tilde{z_i}=\frac{p_ip_a}{p_jp_a+p_ip_a}=1-\tilde{z_j}
\eea
and the momenta of the $m$-parton configuration are obtained by the map
\bea
\tilde{p}_a^\mu=x_{ij,a}\,p_a^\mu\;,\;\;\;
\tilde{p}_{ij}^\mu=p_i^\mu+p_j^\mu-(1-x_{ij,a})p_a^\mu\;.
\eea
Again, four-momentum conservation is trivially fulfilled and the partons remain massless.

The corresponding splitting functions used in Eq.\ (\ref{fidipole}) read
\bea
\langle s|V_{q_ig_j}^a(\tilde{z_i};x_{ij,a})|s^\prime\rangle
&=&
8\pi\mu^{2\epsilon}\alpha_SC_F\left[
\frac2{1-\tilde{z_i}+(1-x_{ij,a})}-(1+\tilde{z_i})-\epsilon(1-\tilde{z_i})
\right]\delta_{ss^\prime}
\;,\nnb\\
\langle \mu|V_{q_i\bar{q}_j}^a(\tilde{z_i};x_{ij,a})|\nu\rangle
&=&
8\pi\mu^{2\epsilon}\alpha_ST_R\left[
-g^{\mu\nu}
-\frac2{p_ip_j}(\tilde{z_i}p_i-\tilde{z_j}p_j)^\mu(\tilde{z_i}p_i-\tilde{z_j}p_j)^\nu\right]
\;,\nnb\\
\langle \mu|V_{g_ig_j}^a(\tilde{z_i};x_{ij,a})|\nu\rangle
&=&
16\pi\mu^{2\epsilon}\alpha_SC_A\left[
-g^{\mu\nu}\left(
\frac1{1-\tilde{z_i}+(1-x_{ij,a})}+\frac1{1-\tilde{z_j}+(1-x_{ij,a})}-2\right)\right.\nnb\\
&&\hspace*{22mm}\left.
+(1-\epsilon)\frac2{p_ip_j}(\tilde{z_i}p_i-\tilde{z_j}p_j)^\mu
                (\tilde{z_i}p_i-\tilde{z_j}p_j)^\nu\right]\;.
\eea

\subsubsection{Initial state emitters with final state spectators}
The next type of dipole function now covers initial state singularities $p_a\!\cdot\!p_i\to0$
with final state spectators, given by
\bea
\label{ifdipole}
\lefteqn{{\cal D}^{ai}_k(p_1,\ldots,p_{m+1};p_a,..)}\nnb\\
&=&
-\frac1{2p_a\!\cdot\!p_i}\frac1{x_{ik,a}}
{\cdot\;}
     {\vphantom{\left\langle\frac{|_|}{|_|}\right\rangle}}_{m,a}\!\!
\left\langle 1,\ldots,\tilde{k},\ldots,m+1;\tilde{ai},..\left|
             \frac{{\bf T}_k\cdot{\bf T}_{ai}}{{\bf T}^2_{ai}}{\bf V}^{ai}_k
             \right|1,\ldots,\tilde{k},\ldots,m+1;\tilde{ai},..\right\rangle_{m,a}\,.\nnb\\
\eea
The parton $\tilde{ai}$, which enters into the $m$-parton matrix element on the r.h.s.\ 
of Eq.\ (\ref{ifdipole}) is given by the splitting of the initial state parton 
$a\to\tilde{ai}+i$.  The relevant kinematic variables in this case are
\bea
x_{ik,a}=1-\frac{p_ip_k}{(p_k+p_i)p_a}\;,\;\;\;
u_i=\frac{p_ip_a}{p_ip_a+p_kp_a}=1-u_k\;,
\eea
and the momenta for the $m$-parton configuration are obtained by
\bea
\tilde{p}_{ai}^\mu=x_{ik,a}\,p_a^\mu\;,\;\;\;
\tilde{p}_{k}^\mu=p_k^\mu+p_i^\mu-(1-x_{ik,a})p_a^\mu\;.
\eea
The splitting matrices ${\bf V}^{ai}_k$ in Eq. (\ref{ifdipole}) are
\bea
\langle s|V^{q_ag_i}_k(u_i;x_{ik,a})|s^\prime\rangle
&=&
8\pi\mu^{2\epsilon}\alpha_SC_F\left[
\frac2{1-x_{ik,a}+u_i}-(1+x_{ik,a})-\epsilon(1-x_{ik,a})
\right]\delta_{ss^\prime}
\;,\nnb\\
\langle s|V^{g_a\bar{q}_i}_k(u_i;x_{ik,a})|s^\prime\rangle
&=&
8\pi\mu^{2\epsilon}\alpha_SC_F\left[
1-\epsilon-2x_{ik,a}(1-x_{ik,a})
\right]\delta_{ss^\prime}
\;,\nnb\\
\langle \mu|V^{q_aq_i}_k(u_i;x_{ik,a})|\nu\rangle
&=&
8\pi\mu^{2\epsilon}\alpha_ST_R\left[
-g^{\mu\nu}x_{ik,a}
+\frac{2u_iu_k}{p_ip_k}\frac{1-x_{ik,a}}{x_{ik,a}}
\left(\frac{p_i}{u_i}-\frac{p_k}{u_k}\right)^\mu
\left(\frac{p_i}{u_i}-\frac{p_k}{u_k}\right)^\nu\right]
\;,\nnb\\
\langle \mu|V^{g_ig_a}_k(u_i;x_{ik,a})|\nu\rangle
&=&
16\pi\mu^{2\epsilon}\alpha_SC_A\left[
-g^{\mu\nu}\left(
\frac1{1-x_{ik,a}+u_i}-1+x_{ik,a}(1-x_{ik,a})\right)\right.\nnb\\
&&\hspace*{22mm}\left.
+(1-\epsilon)+\frac{u_iu_k}{p_ip_k}\frac{1-x_{ik,a}}{x_{ik,a}}
\left(\frac{p_i}{u_i}-\frac{p_k}{u_k}\right)^\mu
\left(\frac{p_i}{u_i}-\frac{p_k}{u_k}\right)^\nu\right]\;.\nnb\\
\eea
The three dipole types discussed up to now (FF, IF, FI) are sufficient to construct the 
subtraction term $\done\sigma^{A}$ for processes with exactly one initial state parton, 
i.e.\ DIS configurations.

\subsubsection{Initial state emitters with initial state spectators}
\label{sii}
The remaining dipole function, only required by processes with two initial state partons, 
covers the case where both, the emitter and the spectator, are initial state particles, 
\bea
\label{iidipole}
\lefteqn{{\cal D}^{ai,b}(p_1,\ldots,p_{m+1};p_a,p_b)}\nnb\\
&=&
-\frac1{2p_a\!\cdot\!p_i}\frac1{x_{i,ab}}
{\cdot\;}
     {\vphantom{\left\langle\frac{|_|}{|_|}\right\rangle}}_{m,ab}\!\!
\left\langle \tilde{1},\ldots,\tilde{m+1};\tilde{ai},b\left|
             \frac{{\bf T}_b\cdot{\bf T}_{ai}}{{\bf T}^2_{ai}}{\bf V}^{ai,b}
             \right|1,\ldots,\tilde{m+1};\tilde{ai},b\right\rangle_{m,ab}\;.\nnb\\
\eea
To describe the splitting, the following kinematic variables are used
\bea
x_{i,ab}=1-\frac{p_ip_a+p_ip_b}{p_ap_b}\;,\;\;\;
\tilde{v}_i=\frac{p_ap_i}{p_ap_b}\;.
\eea
The construction of the $m$-parton kinematics for this dipoles differs from the other 
three cases.  The reason is that in this case the emitter and the spectator are fixed
to remain along the beam axis.  Therefore {\bf all} final state momenta (not only momenta 
of QCD partons) are transformed according to the map
\bea
\tilde{p}_{ai}^\mu=x_{i,ab}\,p_a^\mu\;,\;\;\;
\tilde{p}_{j}^\mu=p_j^\mu-\frac{2p_j\!\cdot\!(K+\tilde{K})}{(K+\tilde{K})^2}(K+\tilde{K})^\mu
+\frac{2p_j\!\cdot\!K}{K^2}\tilde{K}^\mu\;,
\label{iimomtrafo}
\eea
where 
\bea
K^\mu = p_a^\mu+p_b^\mu-p_i^\mu\;\;\;\mbox{\rm and}\;\;\;
\tilde{K}^\mu = \tilde{p}_{ai}^\mu+p_b^\mu\;.
\eea
The momentum of the spectator $p_b$ remains unchanged.  The transformation above can 
also be interpreted as applying a rotation and a boost turning initial state momenta 
back to the beam axis after a mapping similar to the first three cases of dipole 
functions.  Indeed it can be shown that the transformation of final state momenta in 
Eq.\ (\ref{iimomtrafo}) is just a Lorentz transformation. 

However, in this case, the splitting matrices read
\bea
\langle s|V^{q_ag_i,b}(x_{i,ab})|s^\prime\rangle
&=&
8\pi\mu^{2\epsilon}\alpha_SC_F\left[
\frac2{1-x_{i,ab}}-(1+x_{i,ab})-\epsilon(1-x_{i,ab})
\right]\delta_{ss^\prime}
\;,\nnb\\
\langle s|V^{g_a\bar{q}_i,b}(x_{i,ab})|s^\prime\rangle
&=&
8\pi\mu^{2\epsilon}\alpha_ST_R\left[
1-\epsilon-2x_{i,ab}(1-x_{i,ab})
\right]\delta_{ss^\prime}
\;,\nnb\\
\langle \mu|V^{q_aq_i,b}(\tilde{v}_i;x_{ik,a})|\nu\rangle
&=&
8\pi\mu^{2\epsilon}\alpha_SC_F\left[
-g^{\mu\nu}x_{i,ab}
+\frac{2}{\tilde{v}_i\;p_i\!\cdot\!p_b}\frac{1-x_{i,ab}}{x_{i,ab}}
\left(p_i-\tilde{v}_ip_k\right)^\mu
\left(p_i-\tilde{v}_ip_k\right)^\nu\right]
\;,\nnb\\
\langle \mu|V^{g_ig_a,b}(\tilde{v}_i;x_{ik,a})|\nu\rangle
&=&
16\pi\mu^{2\epsilon}\alpha_SC_A\left[
-g^{\mu\nu}\left(
\frac{x_{i,ab}}{1-x_{i,ab}}+x_{i,ab}(1-x_{i,ab})\right)\right.\nnb\\
&&\left.\hspace*{22mm}
+(1-\epsilon)\frac1{\tilde{v}_i\;p_i\!\cdot\!p_b}\frac{1-x_{i,ab}}{x_{i,ab}}
\left(p_i-\tilde{v}_ip_k\right)^\mu
\left(p_i-\tilde{v}_ip_k\right)^\nu\right]\;.\nnb\\
\eea

\subsection{Integrated dipole terms}
\label{finitet}
\subsubsection{Phase space factorisation}
In order to combine the poles of the subtraction function and the virtual matrix 
element the subtraction function has to be integrated analytically over the 
one-parton phase space of the respective splitting.  The rules for the momentum 
mapping from 3 to 2 parton phase spaces have been constructed in Secs.\ 
\ref{sff}-\ref{sii} such that the corresponding phase space exactly factorizes.

As an example, and in order to fix the notation, the case of a final-final dipole, 
${\cal D}_{ij,k}$, will be discussed in the following.  There, the 
three-particle phase space for the partons $i$, $j$ and $k$ (all other partons 
are not affected by the splitting and will be omitted) in $d$ dimensions is 
given by
\bea
\done\phi(p_i,p_j,p_k;Q)
=
\frac{\done^dp_i}{(2\pi)^{d-1}}\delta_+(p_i^2)
\frac{\done^dp_j}{(2\pi)^{d-1}}\delta_+(p_j^2)
\frac{\done^dp_k}{(2\pi)^{d-1}}\delta_+(p_k^2)
(2\pi)^d\delta^{(d)}(Q-p_i-p_j-p_k)\;.\nnb\\
\eea
This can be factorized in terms of the mapped momenta, such that
\bea
\done\phi(p_i,p_j,p_k;Q)
&=&
\done\phi(\tilde{p}_{ij},\tilde{p}_k;Q)
\left[\done p_i(\tilde{p}_{ij},\tilde{p}_k)\right]\;,
\eea 
where $\left[\done p_i(\tilde{p}_{ij},\tilde{p}_k)\right]$, written in terms of 
the kinematic variables defined in section \ref{sff}, reads
\bea
\left[\done p_i(\tilde{p}_{ij},\tilde{p}_k)\right]
&=&
\frac{(2\tilde{p}_{ij}\tilde{p}_k)^{1-\epsilon}}{16\pi^2}
\frac{\done\Omega^{(d-3)}}{(2\pi)^{1-2\epsilon}}\;
\done\tilde{z}_i\;\done y_{ij,k}\;\theta(\tilde{z}_i(1-\tilde{z}_i))
\theta(y_{ij,k}(1-y_{ij,k}))\nnb\\
&&\cdot (\tilde{z}_i(1-\tilde{z}_i))^{-\epsilon}
(1-y_{ij,k})^{1-2\epsilon}y_{ij,k}^{-\epsilon}\;.
\eea
Within the dipole function only the splitting function itself depends on the variables 
$\tilde{z}_i$ and $y_{ij,k}$.  Thus, the integration in $d$ dimensions can be performed 
once and for all, independent of the specific scattering process under consideration.  
The result of the integration for each splitting type can be expanded as a Laurent 
series including double poles ($\sim 1/\epsilon^2$), single poles ($\sim 1/\epsilon$),
and finite terms ($\sim\epsilon^0$).  Further terms of ${\cal O}(\epsilon)$ are unimportant 
here and will be left out.  

All results for the final-final and for all other dipole types can be found in
\cite{Catani:1996vz}.

\subsubsection{Full result}
Having at hand the integrals for each dipole function, all individual dipoles present in
a specific process can be collected to yield the overall infrared divergence of the 
subtraction term.  Then, the starting point for the calculation of jet cross sections 
in the dipole subtraction formalism reads
\bea
\sigma^{\rm NLO} 
=
\sum_{\{m+1\}}\int_{m+1}\left[\done\sigma^{ R}_{\{m+1\}\;|\epsilon=0}-
                              \done\sigma^{ A}_{\{m+1\}\;|\epsilon=0}\right]
+\int_m\left[\sum_{\{m\}}\done\sigma^{ V}_{\{m\}}+
             \sum_{\{m+1\}}\int_1\done\sigma^{ A}_{\{m+1\}}\right]_{\epsilon=0}\;,
\eea
where $\sum_{\{m+1\}}$ denotes the sum over all parton-level processes.  However, the 
important point here is to exactly cancel the poles of the corresponding individual 
one-loop parton-level processes, which is done exclusively for each momentum and flavour
constellation.  Therefore, for each specific $m$-parton process at NLO only a selection 
of dipole functions related to $(m+1)$-parton processes contributes to the cancellation 
of the virtual divergences.  In \cite{Catani:1996vz} it has been shown that this amounts 
to an effective reordering of phase space integrals and sums over parton configurations, 
such that
\bea\label{Eq:fullNLO_reordered}
\sigma^{\rm NLO}
=
\sum_{\{m+1\}}\int_{m+1}\left[\done\sigma^{ R}_{\{m+1\}\;|\epsilon=0}-
                              \done\sigma^{ A}_{\{m+1\}\;|\epsilon=0}\right]
+\sum_{\{m\}}\int_m\left[\done\sigma^{ V}_{\{m\}}+
                         \done\sigma^{\tilde{ A}}_{\{m\}}\right]_{\epsilon=0}\;,
\eea
where $\done\sigma^{\tilde{ A}}_{\{m\}}$ is the integrated dipole term that collects 
the integrals of all dipole functions and thus cancels the singularities of 
$\done\sigma^{ V}_{\{m\}}$.  It is explicitly given by
\bea\label{Eq:BornTimesI}
\done\sigma^{\tilde{ A}}_{\{m\}}
&=&
\left[\done\sigma^{ B}_{\{m\}}\times {\bf I}(\epsilon)\right]\;,
\eea
where $\done\sigma^{ B}_{\{m\}}\times {\bf I}(\epsilon)$ is a shorthand for the 
following procedure:  Write down the expression for $\done\sigma^{ B}_{\{m\}}$, and 
replace the corresponding squared Born-level matrix element
\bea
|{\cal M}_{\{m\}}|^2={}_m\!\langle 1,\ldots,m|1,\ldots,m\rangle_m
\eea
with
\bea
{}_m\!\langle 1,\ldots,m|{\bf I}(\epsilon)|1,\ldots,m\rangle_m\;,
\eea
using the insertion operator ${\bf I}(\epsilon)$ as defined below. 

Finally, the full result for the integrated dipole term and the collinear counterterm as 
defined in Eq.\ (\ref{cct}) for the most general case with hadronic initial states reads
\bea
\label{fullfiniteexcl}
\done\sigma^{\tilde{A}}_{ab}(p_a,p_b)+d\sigma^C_{ab}(p_a,p_b,\mu_F^2)
&=&\;\;\;
\left[\done\sigma^B_{ab}(p_a,p_b)\times{\bf I}(\epsilon)\right]\nnb\\
&&
+\sum_{a^\prime}\int_0^1\done x
      \left[\left({\bf K}^{a,a^\prime}(x)+
                  {\bf P}^{a,a^\prime}(xp_a,x;\mu_F^2)\right)
            \times \done\sigma^B_{a^\prime b}(xp_a,p_b)\right]\nnb\\
&&
+\sum_{b^\prime}\int_0^1\done x
      \left[\left({\bf K}^{b,b^\prime}(x)+
                  {\bf P}^{b,b^\prime}(xp_b,x;\mu_F^2)\right)
            \times \done\sigma^B_{a b^\prime}(p_a,xp_b)\right]\;,\nnb\\
\eea
where $a$ and $b$ again specify the initial state partons.  The summation over $a^\prime$ 
and $b^\prime$ runs over all parton flavours, i.e.\ it includes gluons, quarks and 
anti-quarks occurring in the PDF.

The insertion operator ${\bf I}$ reads
\bea\label{Eq:Insertionoperator}
{\bf I}(\{p\};\epsilon)
=
-\frac{\alpha_S}{2\pi}\frac1{\Gamma(1-\epsilon)}
 \sum_I\frac1{{\bf T}_I^2}{\cal V}_I(\epsilon)
 \sum_{I\neq J}{\bf T}_I\!\cdot\!{\bf T}_J
       \left(\frac{4\pi\mu^2}{2p_Ip_J}\right)^\epsilon\;,
\label{ioperator}
\eea
where the indices $I$ and $J$ run over initial and final state partons.  The universal 
singular functions ${\cal V}_I(\epsilon)$ depend merely on the flavour of $I$ and are
given by
\bea\label{Eq:UniversalV}
{\cal V}_q(\epsilon)
&=&
C_F\left[\frac1{\epsilon^2}+\frac3{2\epsilon}+5-\frac{\pi^2}2+{\cal O}(\epsilon)\right]\nnb\\
{\cal V}_g(\epsilon)
&=&
\frac{C_A}{\epsilon^2}+\left(\frac{11}{6}C_A-\frac23T_RN_f\right)\frac1\epsilon
     +C_A\left(\frac{50}{9}-\frac{\pi^2}{2}\right)-T_RN_f\frac{16}{9}+{\cal O}(\epsilon)\;,
\eea
with $N_f$ being the number of contributing quark flavours. 

The complete singular structure in Eq.\ (\ref{fullfiniteexcl}) is contained in 
$[\done\sigma^B_{ab}(p_a,p_b)\times{\bf I}(\epsilon)]$ and the sum 
$[\done\sigma^B_{ab}(p_a,p_b)\times{\bf I}(\epsilon)]+\done\sigma^V_{ab}(p_a,p_b)$ must be
finite for $\epsilon\to0$.

The finite insertion operators ${\bf K}$ and ${\bf P}$ are given by
\bea\label{Eq:ISinsertion}
{\bf K}^{a,a^\prime}(x)
&=&
\frac{\alpha_S}{2\pi}\left\{\vphantom{\frac{|}{|}}
\bar{K}^{aa^\prime}(x)-K^{aa^\prime}_{\rm F.S.}(x)\right.\nnb\\
&&\hspace*{6mm}\left.
+\delta^{aa^\prime}\sum_i{\bf T}_i\!\cdot\!{\bf T}_a\frac{\gamma_i}{{\bf T}_i^2}
\left[\left(\frac1{1-x}\right)_++\delta(1-x)\right]-
\frac{{\bf T}_b\!\cdot\!{\bf T}_{a^\prime}}{{\bf T}_a^2}\tilde{K}^{a,a^\prime}(x)
\right\}\;,
\eea
and
\bea\label{Eq:Insertion}
{\bf P}^{a,a^\prime}(\{p\};x;\mu_F^2) =
\frac{\alpha_S}{2\pi}P^{a,a^\prime}(x)\frac{1}{{\bf T}_b^2}
\sum_{I\neq b}{\bf T}_I\!\cdot\!{\bf T}_b \ln{\frac{\mu_F^2}{2xp_ap_I}}\;.
\eea
Note that here the index $i$ runs over final state partons only.  The flavour-dependent 
functions $\bar{K}^{aa^\prime}(x)$, $\tilde{K}^{a,a^\prime}(x)$, and $P^{a,a^\prime}(x)$ 
are defined in Appendix \ref{appendix:littlefuncs}.  As already mentioned, the 
factorisation-scheme dependent function $K^{aa^\prime}_{\rm F.S.}(x)$ vanishes in the 
commonly used $\overline{\rm MS}$-scheme.

To obtain the final result for processes with no initial state partons only the 
${\bf I}$-term needs to be considered in Eq.\ (\ref{fullfiniteexcl}).  For processes with 
one initial state parton only, the result is obtained by using the ${\bf I}$-term and one 
of the two integrals over ${\bf K}$ and ${\bf P}$ only, while omitting the contribution 
of $\tilde{K}^{a,a^\prime}(x)$.

\subsection{Freedom in the definition of dipole terms}
\label{dtfreedom}

As stressed before, the singular limits of the dipole functions are fixed by the requirement 
to cancel the singularities of the real correction matrix element.  However, away from this 
limit there is some freedom for modifications.

One possible modification has been presented in \cite{Nagy:2003tz}, where a parameter 
$\alpha$ has been introduced which cuts off a dipole function for phase space regions far 
enough away from the corresponding singularity.  Its main advantage lies in a significant 
reduction of the average number of dipoles terms to be calculated for each phase space 
point of the $(m+1)$-parton phase space of the real correction term.  This constitutes an 
important alleviation of the calculational burden, since the total number of dipole terms 
grows approximately as $m^3$.  The $\alpha$-modified subtraction terms also allow 
nontrivial checks of the implementation, since the total result must be independent 
of $\alpha$.  

The $\alpha$-modified dipole functions have been defined as follows:
\bea
{\cal D}_{ij,k}^\prime&=&{\cal D}_{ij,k}\;\theta(\alpha-y_{ij,k})\;,\nnb\\
{\cal D}_{ij}^{\prime a}&=&{\cal D}_{ij}^a\;\theta(\alpha-1+x_{ij,a})\;,\nnb\\
{\cal D}_{k}^{\prime ai}&=&{\cal D}_{k}^{ai}\;\theta(\alpha-u_{i})\;,\nnb\\
{\cal D}^{\prime ai,b}&=&{\cal D}^{ai,b}\;\theta(\alpha-\tilde{v}_{i})\;.
\label{nagyalpha}
\eea 
They will be employed later, in the implementation presented in this paper.  Of course, 
such a redefinition of the splitting kernels also requires a recalculation of their 
integrals.  The new $\alpha$-dependent insertion operators ${\bf I}$ and ${\bf K}$ have 
been presented in \cite{Nagy:2003tz}.  

Another simple modification is the addition of finite terms to the splitting 
functions, such as
\bea
V_{ij,k}^\prime   &=& V_{ij,k}+y_{ij,k}*C\;,\nnb\\
V_{ij}^{\prime a} &=& V_{ij}^a+(1-x_{ij,a})*C\;,\nnb\\
V_{k}^{\prime ai} &=& V_{k}^{ai}+u_{i}*C\;,\nnb\\
V^{\prime ai,b}   &=& V^{ai,b}+\tilde{v}_{i}*C\;.
\eea
The constant $C$ directly ends up as a finite term in the integral of the splitting 
function and thus it can be easily included in the insertion operators of ${\bf I}$ 
and ${\bf K}$, too.  This again allows checks of the implementation, but it can also be 
employed to improve the numerical behaviour of the phase space integrals and to reduce 
the number of negative events.

\section{Implementation in AMEGIC++}
\noindent
The Catani-Seymour dipole subtraction terms have been implemented in full generality
into the automatic matrix element generator \amegic, based on it's version 2.0
\cite{Amegic20InPreparation}\footnote{
        A brief description of \amegic within the \sherpa framework can be found in 
        \cite{Gleisberg:2003xi}, whereas a full documentation of the (partly obsolete) 
        version 1.0 is given in \cite{Krauss:2001iv} with some extensions and results
        discussed in \cite{Gleisberg:2003ue}-\cite{Hagiwara:2005wg}.  An update on 
        the helicity formalism as it is used in the current version is documented 
        in \cite{Gleisberg:2003ue}.}.  
In particular this translates into \amegic being able to automatically generate 
all relevant parts of the NLO matrix element within the subtraction method except for 
the virtual matrix element.  It can be applied to any process with massless partons 
for which the real correction ME can be generated, an extension to allow also for
massive particles is foreseen. This includes standard model processes as well as 
implemented extensions, as long as there are no new strongly interacting particles involved.  
For standard model processes the boundary is currently at about six-eight partons 
(initial and final state).

The new implementation aimed at a maximal reuse of already developed automated methods
of amplitude generation and process management.  Therefore, first a brief overview over 
the relevant parts of the code are given before the new implementation is described in
some detail.

\subsection{ME generation}
\subsubsection{Amplitudes}

The matrix element evaluation in the $C++$-code \amegic is based on the evaluation of 
Feynman amplitudes using a helicity method based on the developments in
\cite{De Causmaecker:1981by}-\cite{Ballestrero:1992dv}.  The fundamental idea of this 
method is to introduce a helicity basis, in terms of which all physical spinors can be 
expressed.  This allows to compute each amplitude directly as a complex function of 
physical momenta and helicity/spin states instead of computing traces of spinor products 
and $\gamma$-matrices for squared amplitudes.   The colour within any amplitude is 
treated separately, i.e.\ in the first step all colour factors (the SU(3) structure 
constants ${f}^{abc}$ and ${t}^a_{ij}$) corresponding to the $k$-th amplitude $A_k$ are 
collected in an array ${\bf C}_k$.

Squared matrix elements can thus be written as
\bea
|M|^2&=&\sum_{i,j} \left[\left(A_iA_j^*\right)
                         \left({\bf C}_i\!\cdot\!{\bf C}_j^\dagger\right)\right]\;,
\eea
and hence a colour matrix of complex numbers
\bea
c_{i,j}&=&{\bf C}_i\!\cdot\!{\bf C}_j^\dagger
\eea
is generated once for each process.  In an initialisation run, \amegic generates all 
formulae for the amplitude calculation of the user-specified parton level processes 
including the colour matrix $c_{i,j}$ (using a set of replacement rules for the colour 
algebra), it simplifies the expressions for the helicity amplitudes by identifying and 
factoring out common pieces and finally stores everything in C++ libraries. 

\subsubsection{Process structure and organisation}
\label{amegicpo}
Since typically many parton level processes contribute to jet cross section calculations,
usually a long list of processes needs to be computed. The corresponding structure in 
\amegic is as follows:
\begin{itemize}
\item Any parton level process is represented by the class {\tt Single\_Process},
\item {\tt Process\_Group} contains a (possibly recursive) list of such processes or
        groups of processes.
\end{itemize}
All parton level processes sharing a specific common set of properties are grouped together 
in two or three levels of groups.  In many cases there are subprocesses contributing to the
same jet cross section which are very similar.  Therefore \amegic applies a procedure to 
identify such processes in order to save computer resources and accelerate the calculation.
The following checks are performed:
\begin{itemize}
\item Direct comparison of amplitudes: check for processes that have identical graphs, where
        all involved particles have the same masses, widths and underly the same interactions
        (with coupling constants that differ at most in a constant factor).\\ 
        Example: QCD processes that differ in quark flavours only.
\item Numerical comparisons: check if the numerical result for a squared matrix element at 
        a given phase space point is the same.\\
        Example: a quark is replaced by an anti-quark w.r.t. to the other process.
\end{itemize}
For a set of processes that can be identified by this it is enough to compute one to know 
them all.  In such a case, the corresponding matrix element squared is calculated only
once and then recycled by the other processes.

\subsection{Generation of CS dipole terms}
\subsubsection{Colour and spin correlations}
The starting point of the Catani-Seymour algorithm is detailed in Eqs.\ 
(\ref{Eq:fullNLO_reordered}) and (\ref{Eq:BornTimesI}), supplemented with expressions 
like the one in Eq.\ (\ref{Eq:OneDipoleSubtractionTerm}) for the individual dipole 
subtraction terms.  The latter states that for any given process the Catani-Seymour 
dipole subtraction term for the real $(m+1)$-parton correction term consists of the 
corresponding $m$-parton matrix element at Born level plus an additional operator that 
acts on colour and spin space.  For the latter, only the limit $\epsilon\to0$ needs to 
be considered.
\begin{itemize}
\item Colour operator:\\ 
        In all four dipoles, Eq. (\ref{ffdipole}), (\ref{fidipole}), (\ref{ifdipole}), and 
        (\ref{iidipole}) colour-correlated tree-amplitudes of the form
        \bea
        |M_m^{i,k}|^2&=&{}_m\!\!\langle 1,\ldots,m|{\bf T}_i\cdot{\bf T}_k|1,\ldots,m\rangle_m
        \eea
        occur, where $i$ labels the emitter and $k$ the spectator. Denoting the colour indices
        of the external legs of the tree process explicitly by $a_j$ and $b_j$, this can be
        cast into
        \bea
        \lefteqn{|M_m^{i,k}|^2}\nnb\\&=&
        {}_m\!\!\langle 1^{a_1}\ldots i^{a_i}\ldots k^{a_k}\ldots m^{a_m}|
        \delta_{a_1b_1}\ldots T^c_{a_ib_i}\ldots T^c_{a_kb_k}\ldots \delta_{a_mb_m}
        |1^{b_1}\ldots i^{b_i}\ldots k^{b_k}\ldots m^{b_m}\rangle_m\;,\nnb\\
        \eea
        where $T^c_{ab}=if^{acb}$, if the associated particle is a gluon, and 
        $T^c_{ij}=t^c_{ij}$, if the associated particle is a quark.  In other words, the 
        colour structure for dipole terms can be generated by adding a gluon connecting
        the emitter with the spectator as illustrated in Fig.\ \ref{dipolecolor}. 
        \begin{figure}
        \begin{center}
        \includegraphics[width=100mm]{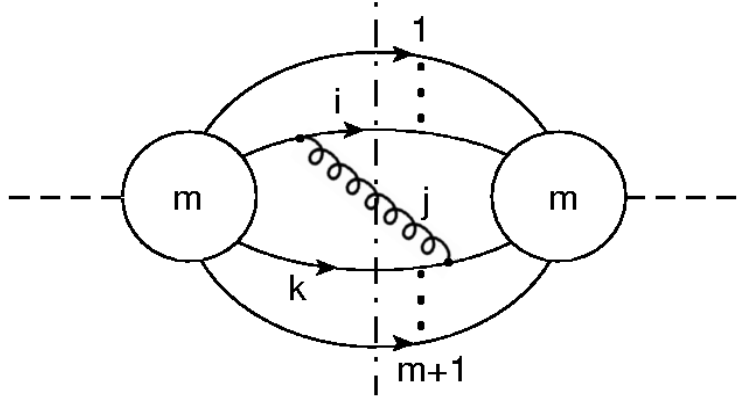}
        \caption{\label{dipolecolor}Construction of the colour matrix for dipole terms: 
                        a gluon connects emitter and the spectator.}
        \end{center}
        \end{figure}
        The colour matrix for a dipole term is recomputed after this insertion using
        the available evaluation tool in \amegic.
\item Spin space:\\ 
        For a quark splitting all spin-matrices are just proportional to 
        $\delta_{ss^\prime}$, translating the quark spin to be exactly the same as for 
        the Born-level $m$-parton matrix element. 

        For the case of a gluon splitting, however, there are non-trivial correlation 
        matrices.  All of them can be cast into the generic form
        \bea
        V^{\mu\nu}=\langle\mu|V|\nu\rangle
        &\propto&-g^{\mu\nu}+\frac{\tilde{p}^\mu\tilde{p}^\nu}{B\,\tilde{p}^2}\;,
        \label{pol_spindef}
        \eea
        where $B$ and $\tilde{p}$ are functions of the kinematic variables and momenta
        of the corresponding splitting. Their values are listed in Table 
        \ref{dipolespincorr}.
        \begin{table}
        \begin{center}
        \begin{tabular}{llll}
                dipole type & splitting:  & $\tilde{p}^\mu$ & $B$ \\ \hline\hline
                \multirow{2}{*}{FF} & $g\to q\bar{q}$ & 
                        \multirow{2}{*}{$\tilde{z_i}p_i^\mu-\tilde{z_j}p_j^\mu$} & 
                                        $1/(4\tilde{z_i}\tilde{z_j})$ \\[1mm]
                & $g\to gg$  &  & 
                        $\left(2-\frac1{1-\tilde{z_i}(1-y_{ij,k})}-
                                \frac1{1-\tilde{z_j}(1-y_{ij,k})}\right)/
                                (2\tilde{z_i}\tilde{z_j})$ \\ \hline
                \multirow{2}{*}{FI} & $g\to q\bar{q}$ &  
                        \multirow{2}{*}{$\tilde{z_i}p_i^\mu-\tilde{z_j}p_j^\mu$} & 
                                        $1/(4\tilde{z_i}\tilde{z_j})$   \\[1mm]
                & $g\to gg$  &  &  
                        $\left(2-\frac1{1-\tilde{z_i}+(1-x_{ij,a})}-
                                \frac1{1-\tilde{z_j}+(1-x_{ij,a})}\right)/
                                (2\tilde{z_i}\tilde{z_j})$ \\ \hline
                \multirow{2}{*}{IF} & $g\to q\bar{q}$ &  
                        \multirow{2}{*}{$\displaystyle\frac{p_i^\mu}{u_i}-
                                        \frac{p_k^\mu}{u_k}$} & 
                                        $-\frac14x_{ik,a}^2/(1-x_{ik,a})$  \\[1mm]
                & $g\to gg$  &  & 
                        $\frac12\left(1-\frac{1}{1.-x_{ik,a}+u_i}-x_{ik,a}(1-x_{ik,a})\right)
                        x_{ik,a}/(1-x_{ik,a})$ \\ \hline
                \multirow{2}{*}{II} & $g\to q\bar{q}$ &  
                        \multirow{2}{*}{$p_i^\mu-\tilde{v}_ip_k^\mu$} & 
                                        $-\frac14x_{i,ab}^2/(1-x_{i,ab})$ \\[1mm]
                & $g\to gg$  &  & 
                        $-\frac12\left(\frac{1}{1.-x_{ik,a}}+(1-x_{ik,a})\right)x_{ik,a}^2/
                        (1-x_{ik,a})$ \\ \hline
        \end{tabular}
        \parbox[c]{16.5cm}{\caption{\label{dipolespincorr} Values for the functions 
                        defined in Eq.\ (\ref{pol_spindef}).  The variables are defined 
                        in the corresponding sections \ref{sff}-\ref{sii}.  The dipole type 
                        FF refers to the case where emitter and spectator are final state 
                        partons, IF refers to the case where the emitter is an initial state 
                        parton and the spectator a final state parton, etc..}}
        \end{center}
        \end{table}
\end{itemize}
The structure of the splitting tensor as given in Eq.\ (\ref{pol_spindef}) is very similar to 
the polarisation sum for massive vector bosons in unitary gauge, except for the factor $B$ 
and the fact that $\tilde{p}$ can be timelike or spacelike. This analogy can be used to 
replace the tensor by a polarisation sum, i.e.\
\bea
-g^{\mu\nu}+\frac{\tilde{p}^\mu\tilde{p}^\nu}{B\,\tilde{p}^2}
&=&
\sum_{\lambda}\xi^\lambda\epsilon_\lambda^\mu(\tilde{p},B)
        \left(\epsilon_\lambda^\nu(\tilde{p},B)\right)^*\;.
\eea
Here the summation index $\lambda$ runs over four values, $+$, $-$, $l$ and $s$.  
$\xi^\lambda$ is a sign that cannot be absorbed into the polarisation vectors 
$\epsilon_\lambda$.  For a gauge boson with momentum
\bea
\tilde{p}^\mu&=&\left(\tilde{p}_0,|\tilde{\vec{p}}|\sin{\theta}\cos{\phi},
|\tilde{\vec{p}}|\sin{\theta}\sin{\phi},|\tilde{\vec{p}}|\cos{\theta}\right)\;,
\eea
the polarisation vectors are defined as 
\bea
\epsilon_\pm^\mu
&=&
\frac1{\sqrt{2}}\left(0,\cos{\theta}\cos{\phi}\mp i\sin{\phi},
                      \cos{\theta}\sin{\phi}\pm i\cos{\phi},-\sin{\theta}\right)\;,\nnb\\
\epsilon_l^\mu
&=&
\frac1{\tilde{p}^2}\left(|\tilde{\vec{p}}|,
                        \tilde{p}_0\frac{\tilde{\vec{p}}}{|\tilde{\vec{p}}|}\right)\;,\nnb\\
\epsilon_s^\mu
&=&
\sqrt{\frac{1-B}{B\;,\tilde{p}^2}}\;\tilde{p}^\mu\;,
\eea
and the sign factors are given by
\bea
\xi^\pm = 1\;,\;\;\;\;
\xi^l = \left\{\begin{matrix}
                        -1\;\; {\rm if}\;\; \tilde{p}^2<0\\ 
                        +1\;\; {\rm if}\;\; \tilde{p}^2>0
                \end{matrix}\right.\;,\;\;\;\;
\xi^s=\left\{\begin{matrix}
                        +1\;\; {\rm if}\;\; \tilde{p}^2<0\hphantom{\;;\;\;B>1}\\ 
                        -1\;\; {\rm if}\;\; \tilde{p}^2>0\;;\;\;B>1\\
                        +1\;\; {\rm if}\;\; \tilde{p}^2>0\;;\;\;B<1
                \end{matrix}\right.\;.
\eea
In order to calculate the dipole matrix element, the polarisation vectors of the splitting 
gluon are then replaced by the ones defined above. 

\subsubsection{Organisation and process management}
To construct all dipole functions necessary to cancel the infrared divergencies of a given
parton level real-correction process firstly all pairs of partons have to be determined that 
might emerge from the splitting of an emitter parton (initial state partons are charge 
conjugated for this procedure).  This might be any quark (or anti-quark) and a gluon, two 
gluons or a quark and an anti-quark of the same flavour.  Secondly, each of those pairs is 
combined with any possible third parton (acting as spectator) to define all possible dipole 
functions.

Any individual dipole function is thus specified by:
\begin{enumerate} 
\item type (the specific combination of initial and final state for emitter and spectator),
\item the specific flavours involved in the splitting, and
\item the corresponding $m$-parton matrix element and its emitter and spectator particles.
\end{enumerate}

In order to construct the individual dipole functions, given by
\bea
{\cal D} &=& A_i\,C^\prime_{ij}\,A_j^*\, F(...)\;.
\eea
the following ingredients are necessary:
\begin{enumerate}
\item A rule to map the $(m+1)$-parton phase space onto an $m$-parton phase space.
\item The corresponding splitting function for the dipole.  This consists of two parts, a 
        scalar function $F(...)$ of the kinematic variables of the splitting and a spin
        correlation matrix.  As discussed above, for quark splittings the matrix is simply 
        $\delta_{ss^\prime}$, for gluon splitting the matrix is represented by an outer 
        product of pseudo-polarisation vectors, which are also functions of the kinematic 
        variables of the splitting.
\item The colour matrix $C^\prime_{ij}$, respecting the extra colour correlation.
\item Amplitudes $A_i$ of the corresponding $m$-parton matrix elements. For gluon splitting 
        cases these amplitudes have to be calculated replacing polarisation vectors of the 
        splitting gluon by the pseudo-polarisation vectors introduced above.
\end{enumerate}

The calculation of any dipole function is organized in the class {\tt Single\_DipoleTerm},
each instance of this class representing one dipole.  This class controls the ingredients 
for the calculation: Firstly there is a Born-level $m$-parton matrix element of the original
\amegic implementation, just extended such that it includes the additional colour correlation.  
Secondly there is a class {\tt Dipole\_Splitting\_Base} that completely organizes the 
splitting function itself.  Specified by the type of the dipole (initial and final states 
for emitter and spectator) and the type of the splitting (determined by the contributing 
flavours) it takes care of the mapping between the $m+1$-parton and the $m$-parton phase 
spaces and of the calculation of the splitting function (including the polarisation vectors 
to encode the spin correlation).

Above that the class {\tt Single\_Real\_Correction} handles all contributions to an infrared 
regularized parton level process.  This consists firstly of an $(m+1)$-parton tree level 
matrix element in the original \amegic implementation. Secondly it contains a list of single 
dipole functions, simply determined by looping over all partons and selecting valid dipole 
configurations.  The classes {\tt Single\_Real\_Correction} and {\tt Single\_Process} are 
derived from a common base class in a way such that the class {\tt Process\_Group} can be 
reused to also organize the infrared regularized parton level process in groups of common 
features up to all subprocesses contributing to a jet cross section.

Similarly to the case of tree level processes in \amegic, also here a mapping of parton level 
processes that lead to identical or proportional results can be used to speed up the 
calculation and save computer resources.  To this end, the following automatic identification 
strategies are implemented:
\begin{itemize}
\item If two real correction processes can be mapped (using strategies described in section 
        \ref{amegicpo}) then also the whole {\tt Single\_Real\_Correction} is mapped. 
\item For single dipole terms a unique identification algorithm proceeds as follows: 
        Two terms can be mapped if the included $m$-parton process can be mapped and if 
        the three particle labels (numbering the the external particles of the real 
        correction process) to identify a dipole are identical.
\item Many of the born matrix elements within the dipole terms will be identical. However,
        since different dipoles require different momentum mappings they have to be 
        recalculated.  Only the calculation routine can be shared.
\end{itemize}

\subsection{Generation of the finite part of integrated dipole terms}
\subsubsection{Analytical structure of the full result}
\noindent
The starting point of the discussion of the finite pieces of the integrated dipole terms is 
Eq.\ (\ref{fullfiniteexcl}), where now the phase space integration as well as the summation
and integration over the incoming parton flavours and momenta is made explicit.  Then, terms 
inside the $m$-parton integral come from subtraction terms integrated over the phase space 
of the extra parton emission and from the collinear counterterm for the general case of a 
NLO cross section with initial state partons.  The terms inside the $(m+1)$-parton phase
space integral in contrast corresponds to the dipole subtraction bit.  Altogether, and   
including the convolution with parton distribution, the relevant term to be evaluated
can thus be cast into
\bea\label{fullfinite}
\lefteqn{
\sum_{a,b} \int \done\eta_1\done\eta_2 f_a(\eta_1,\mu_F^2)f_b(\eta_2,\mu_F^2)
        \left\{\int_{m+1}\done\sigma^A_{ab}(\eta_1p,\eta_2\bar p)+
               \int_{m}\done\sigma^C_{ab}(\eta_1p,\eta_2\bar p,\mu_F^2)\right\}}\nnb\\
&=&\sum_{a,b} \int\done\eta_1\done\eta_2 f_a(\eta_1,\mu_F^2)f_b(\eta_2,\mu_F^2)
\left\{\int_m\left[\done\sigma^B_{ab}(\eta_1p,\eta_2\bar p)\times{\bf I}(\epsilon)\right]
        \right.\nnb\\
&&\hspace*{3cm}
      +\sum_{a^\prime}\int_0^1dx\int_m
                \left[\left({\bf K}^{a,a^\prime}(x)
                           +{\bf P}^{a,a^\prime}(x\eta_1p,x;\mu_F^2)\right)
                      \times\done\sigma^B_{a^\prime b}(x\eta_1p,\eta_2\bar p)\right]\nnb\\
&&\hspace*{3cm}\left.
        +\sum_{b^\prime}\int_0^1dx\int_m
                \left[\left({\bf K}^{b,b^\prime}(x)
                           +{\bf P}^{b,b^\prime}(x\eta_2\bar p,x;\mu_F^2)\right)
                        \times\done\sigma^B_{a b^\prime}(\eta_1p,x\eta_2\bar p)\right]\right\}.
\nnb\\
\eea 
The only correlation of the insertion operators ${\bf I}$, ${\bf P}$, and ${\bf K}$ with the 
Born level matrix element is within color space.  To be more specific, this implies that
only the following structures emerge
\bea
\done\sigma^B_{ab}(p_a,p_b)
&=&
\hphantom{1}_m\langle 1,\ldots,m;a,b||1,\ldots,m;a,b\rangle_m\;\;{\rm and}\nnb\\
\done\sigma^{B(i,j)}_{ab}(p_a,p_b)
&=&
\hphantom{1}_m\langle 1,\ldots,m;a,b|{\bf T}_i\cdot{\bf T}_j|1,\ldots,m;a,b\rangle_m
\label{ccxs}
\eea
for all $i\neq j$, where $i$ and $j$ may label both final and initial state partons.  Since 
any of the appearing matrix elements with insertion operators can be written as a sum of 
such structures, the color factors will be skipped in the following and the operators 
will be treated simply as scalar functions.  

The terms ${\bf P}$ and ${\bf K}$ induce dependences on $x$, which combined yield
result in the structure
\bea
\left(g(x)\right)_+ + \delta(1-x)h(x) + k(x)\;.
\label{acontrib}
\eea
Here, $h(x)$ and $k(x)$ are regular functions in $x$ and the '+'-distribution is defined
by its action on a generic test function $a(x)$
\bea
\int_0^1 dx\, a(x) \left(g(x)\right)_+ = \int_0^1 dx \left[a(x)-a(1)\right] g(x)\,.
\eea
Then the r.h.s.\ of Eq.\ (\ref{fullfinite}) can be cast into the form
\bea
\lefteqn{\sum_{a,b} \int\done\eta_1\done\eta_2 f_a(\eta_1,\mu_F^2)f_b(\eta_2,\mu_F^2)}\nnb\\
&&\int_m\left\{{\bf I}(\epsilon)\done\sigma^B_{ab}(\eta_1p,\eta_2\bar p)
        \vphantom{\sum_{b^\prime}}\right.\nnb\\
&&+\sum_{a^\prime}\left[\int_0^1\done x\left(g^{a,a^\prime}(x)\left[
        \done\sigma^B_{a^\prime b}(x\eta_1p,\eta_2\bar p)-
        \done\sigma^B_{a^\prime b}(\eta_1p,\eta_2\bar p)\right]
        +k^{a,a^\prime}(x) \done\sigma^B_{a^\prime b}(x\eta_1p,\eta_2\bar p)\right)\right.\nnb\\
&&\hspace*{10mm}
        +h^{a,a^\prime}(1) \done\sigma^B_{a^\prime b}(\eta_1p,\eta_2\bar p)
        \left.\vphantom{\int_0^1}\right]\hphantom{a}\nnb\\
&&+\sum_{b^\prime}\left[\int_0^1\done x\left(g^{b,b^\prime}(x)\left[
        \done\sigma^B_{a b^\prime}(\eta_1p,x\eta_2\bar p)-
        \done\sigma^B_{a b^\prime}(\eta_1p,\eta_2\bar p)\right]
        +k^{b,b^\prime}(x) \done\sigma^B_{a b^\prime}(\eta_1p,x\eta_2\bar p)\right)\right.\nnb\\
&&\hspace*{10mm}
        +h^{b,b^\prime}(1) d\sigma^B_{a b^\prime}(\eta_1p,\eta_2\bar p)
        \left.\left.\vphantom{\int_0^1}\right]\vphantom{\sum_{b^\prime}}\right\}.
\eea
The functions $g^{a,a'}(x)$, $k^{a,a'}(x)$, and $h^{a,a'}(1)$ can be read off the
corresponding functions in App.\ \ref{appendix:littlefuncs}.  

Computationally the most demanding part is the actual Born-level cross section 
$\done\sigma^B_{ab}$, due to its potentially expensive multi-particle matrix 
element, which typically suffers from factorial growth with the number of external particles.  
Thus, the calculation can be significantly accelerated if the expression is rearranged such 
that $\done\sigma^B_{ab}$ has to be computed only once for a single configuration at a given 
phase space point.  This can be achieved by changing the integration variables $\eta$ to 
$\eta^\prime=x\eta$. After renaming $\eta^\prime$ back to $\eta$ and reordering the summation 
over $a$ and $a^\prime$ ($b$ and $b^\prime$) the expression above reads
\bea
\lefteqn{\sum_{a,b} \int d\eta_1d\eta_2 f_a(\eta_1,\mu_F^2)f_b(\eta_2,\mu_F^2)
        \int_m\done\sigma^B_{ab}(\eta_1p,\eta_2\bar p)
        \times \left\{{\bf I}(\epsilon)\vphantom{\sum_{b^\prime}}\right.}\nnb\\
&&+\sum_{a^\prime}\int_{\eta_1}^1\done x
        \left[\frac{f_{a^\prime}(\eta_1/x,\mu_F^2)}{x\,f_a(\eta_1,\mu_F^2)}
                \left(g^{a^\prime\!,a}(x)+k^{a^\prime\!,a}(x)\right)-
              \frac{f_{a^\prime}(\eta_1,\mu_F^2)}{f_a(\eta_1,\mu_F^2)}
                g^{a^\prime\!,a}(x)\right]\nnb\\
&&+\sum_{a^\prime}\frac{f_{a^\prime}(\eta_1,\mu_F^2)}{f_a(\eta_1,\mu_F^2)}
        \left(h^{a^\prime\!,a}-G^{a^\prime\!,a}(\eta_1)\right)\nnb\\
&&+\sum_{b^\prime}\int_{\eta_2}^1\done x
        \left[\frac{f_{b^\prime}(\eta_2/x,\mu_F^2)}{x\,f_b(\eta_2,\mu_F^2)}
                \left(g^{b^\prime,b}(x)+k^{b^\prime,b}(x)\right)-
              \frac{f_{b^\prime}(\eta_2,\mu_F^2)}{f_b(\eta_2,\mu_F^2)}
                g^{b^\prime,b}(x)\right]\nnb\\
&&+\sum_{b^\prime}\frac{f_{b^\prime}(\eta_2,\mu_F^2)}{f_b(\eta_2,\mu_F^2)}
        \left(h^{b^\prime,b}-G^{b^\prime,b}(\eta_2)\right)
        \left.\vphantom{\sum_{b^\prime}}\right\},
\label{fullfinitreorderd}
\eea
where the $G^{a,b}(\eta)=\int_0^\eta dx\, g^{a,b}(x)$ are analytically computed.

The insertion operator ${\bf I}(\epsilon)$, Eq.\ (\ref{Eq:Insertionoperator}) is given as 
a Laurent series in $\epsilon$.  For the implementation the interesting part is 
$\propto\epsilon^0$, since the poles must have been analytically extracted before\footnote{ 
        For testing purposes, however, it is trivial to also determine the coefficients 
        of the $\epsilon^{-2}$- and $\epsilon^{-1}$-poles and to compare with known
        results of virtual correction terms.}.

\subsubsection{Implementation and Organisation}
The numerical calculation of the finite contributions from integrated counterterms is 
organized as Eq.\ (\ref{fullfinitreorderd}) suggests, i.e.\ the basic unit 
(class {\tt Single\_Virtual\_Correction}) covers everything that is associated 
with a specific $m$-parton cross section. 

For the actual calculation, basically all colour correlated matrix elements in 
Eq.\ (\ref{ccxs}) are necessary. The contributing amplitudes are, of course, the same for 
all of them, only the colour matrix is different.  Therefore, a generalized version of
{\tt Single\_Process} is employed that is able to deal with a multitude of colour matrices
to calculate all required matrix elements at once.  Anything else needed for the calculation
of the finite contribution is a long list of rather simple scalar functions and constants.  
The integration over $x$ is done numerically, i.e.\ for each set of external momenta
$x$ is diced within the corresponding interval.

\subsection{Phase space integration}
Together with the automatic generation of matrix elements \amegic also generates specific,
process-dependent phase-space mappings for efficient integration.  There, some 
{\it a priori} knowledge about the integrand is used 
\cite{Berends:1994xn}-\cite{van Hameren:2002tc} together with self-adaptive Monte Carlo 
integration methods \cite{Maltoni:2002qb,Kleiss:1994qy}-\cite{Ohl:1998jn}.  Here, the general 
method will briefly be summarized for the case of tree-level processes and its application 
to the integrals coming with the subtraction method will be discussed.

\subsubsection{Importance sampling and Multi-channel integration}

The general idea behind importance sampling is to improve the numerical behaviour of an 
integrand by a change of integration variables,
\bea
\int f(x)dx=\int \frac{f(x(y))}{g(x(y))}dy\;,\;\;{\rm where}\;\;\;\;
\frac1g=\frac{dx(y)}{dy}\;.
\eea
The new variable $y$ is chosen in a way such that $\frac{f}{g}$ is a sufficiently 
smooth function, leading to a reduced error estimate of the integration.  Typically, 
the weight $g$ is chosen as a simplification/approximation of $f$, such that the 
integral $y=\int gdx$ can be analytically solved.
 
For phase space integrals maps relating vectors of uniformly distributed random 
numbers $\{a_i\}$ inside the interval $[0,1]$ to the four-momenta of the external 
particles of a physical process $\{p_j\}$
\bea
\{p_j\}&=&X(\{a_i\})
\eea
are in the center of the sampling process.  The weight function in such a case $g$ 
is then determined by
\bea
\frac1g &=& \frac{\done\Phi_n(X(\{a_i\}))}{\done\{a_i\}}\,.
\eea
For a single squared amplitude it is easy to determine suitable momentum mappings and 
weights: invariant masses of the propagators are determined according to the propagator 
and angles for particle splitting are chosen isotropically; the finial state momenta 
are then determined out of those variables.  For instance, for a massless propagator 
the invariant $s$ would be generated by
\bea
\label{psms}
s&=&\left[a s_{\rm max}^{1-\nu}+(1-a) s_{\rm min}^{1-\nu}\right]^{\frac1{1-\nu}}\;,
\eea
with the corresponding weight 
\bea
g&=&\frac{1-\nu}{s_{\rm max}^{1-\nu}-s_{\rm min}^{1-\nu}}\frac1{s^\nu}\;.
\eea
The constants $s_{\rm max}$ and $s_{\rm min}$ are upper and lower boundaries of the 
invariant mass, which depend on the overall topology of the phase space point and 
potential cuts; $\nu$ in contrast is an effective exponent for the propagator, subject 
to choice.

Weight distributions for contributions from several amplitudes can be then combined 
using the multi-channel method.  A total weight function $G$ is defined through
\bea
G&=&\sum_k \alpha_k\; g_k\;,
\eea
where the $g_k$ are the weight functions for the single contributions (channels) and 
the $\alpha_k$ are arbitrary coefficients with $\alpha_k>0$ and $\sum_k \alpha_k=1$.  
The corresponding momentum mapping is then given by
\bea
{\bf X}(\{a_i\},\tilde\alpha)&=&X_k(\{a_i\})\;,\;\;{\rm for}\;\;
\sum_{l=1}^{k-1}\alpha_l<\tilde\alpha<\sum_{l=1}^{k}\alpha_l\;.
\eea
The multi-channel method relies on automatically adapting the coefficients $\alpha_k$ 
such that the variance of the phase space integral is minimized.

{\bf Further refinement}

The efficiency of the integrator is improved if additionally the self-adaptive \vegas 
algorithm \cite{Lepage:1980dq} is applied on the channels. \vegas is very efficient 
in the numerical adaptation to functions, where the peaking behaviour is not too extreme 
and which are factorisable to a product of one-dimensional functions.  This is clearly 
not given for full matrix elements. However, the structure represented by a single 
channel fulfills this condition.  Thus, in \amegic \vegas is used to adapt selected 
channels to structures that go beyond their rough approximations and which are typically 
hard to specify analytically or which are a priori unknown. 

For each channel \vegas is used to generate a mapping $\xi$ from uniformly distributed 
random numbers to a non-uniform distribution, still inside the interval $[0,1]$, and 
a corresponding weight $v_k$.  To combine this with the multi-channel method the 
mapping $X(\{a_i\})$ for single channels must meet the requirement to be invertible.  
The full map reads
\bea
{\bf X}(\{a_i\},\tilde\alpha)&=&X_k(\xi_k(\{a_i\}))\;,\;\;{\rm for}\;\;
\sum_{l=1}^{k-1}\alpha_l<\tilde\alpha<\sum_{l=1}^{k}\alpha_l\;.
\eea
For a momentum configuration $\{p_j\}$ the weight is therefore given by
\bea
G(\{p_j\})&=&\sum_k \alpha_k\; g_k(\{p_j\})\; v_k(X_k^{-1}(\{p_j\}))\;.
\eea

{\bf Subtracted processes}

The subtraction method necessitates the evaluation of two independent integrals, namely 
integrals over the $m$-parton and the $(m+1)$-parton phase space.  In both cases mappings 
generated for the tree level process of the same dimensionality are used.

For the integration of the $(m+1)$-parton phase space soft and collinear regions must be 
included.  In this case the lower limit for the invariant masses of many propagators 
(e.g.\ Eq.\ (\ref{psms})) must be zero.  To keep the integral over the weight finite the 
exponent $\nu$ must be set to a number smaller than 1.  The actual shape of those 
propagators is hard to specify {\it a priori}.  It depends on the jet definition and 
on the balance between the real correction process and the subtraction term (the integrand 
can be positive or negative).  Taken together, however, it seems not unreasonable to 
assume a small exponent.  The \vegas refinement adapts very good to the actual shape 
and the final integration efficiency after optimisation has only a weak dependence on 
the initial value of $\nu$.  Since the \vegas algorithm optimizes on the variance of 
the integrand it can, to some extend, also deal with the numerical problems related to 
``missed binning'', which will be discussed in the following section.

The $m$-parton phase space is much simpler. Since most parts of the integrand are 
proportional to the born matrix element it tends to work very well with this phase 
space setup.

\subsection{Cuts and analysis framework for NLO calculations}
Triggers and observables for NLO calculations have to be chosen with care.  The general 
strict requirement not to spoil the cancelation of infrared divergencies has already been 
discussed in section \ref{nloobservable}.

Before going into any details concerning cuts, it is important to notice that a rule is 
mandatory of how cuts act on the different contributions to the NLO cross section.  This 
rule must exist in a $m$-parton and a $(m+1)$-parton version, where the latter needs to 
satisfy the conditions of infrared safety in degenerate phase space regions.  In practical 
terms, this implies that the $(m+1)$-parton version of the cut must allow for exactly 
one parton to become soft or collinear, while the $m$-parton version has to omit all 
singular regions.
 
Second, Eq.\ (\ref{observablereq}) requires for the cut of the $m+1$ phase space integral 
to be applied separately to the real correction process (using the $m+1$-parton version) 
and to each dipole term (using the $m$-parton version, applied on the momenta of the mapped 
$m$-parton configuration).  In general there might be kinematic configurations, where the 
real correction process ends up outside the accepted phase space region but some dipole 
terms do not and vice versa.  This leads to the problem of ``missed binning'': if such a 
configuration occurs close to a singular region, large contributions result, which do not 
cancel completely.  Ultimately, this leads to large numerical fluctuations, which need
to be addressed.  This is a common issue for all subtraction methods.

So far, the following cuts have been made available in \amegic:
\begin{itemize}
\item A simple cut for jets is implemented as follows: a suitable jet algorithm 
        (e.g. $k_T$) \cite{Catani:1991hj}-\cite{Blazey:2000qt} is used to construct jets 
        from the final state partons and their momenta.  Then the number of jets above a 
        given $p_T$-cut is counted.  A phase space point is valid if this number is 
        greater or equal $m$.
\item Of course also cuts that only act on particles not taking part in strong interactions 
        can be applied.  If initial-initial dipoles are present this also has to be done 
        separately for the real correction and for the dipole terms, since the momentum 
        mapping in this case modifies all final state particle momenta.  
        Implemented are cuts on invariant masses, on total or transverse energies, 
        on rapidities or on particle angles w.r.t. the beam.  
\end{itemize}

Sherpa's \analysis-package has been extended to be able to deal with weighted events from 
the NLO subtraction procedure.  For example, and to be more specific, consider the case of 
a cross section which is differential to some infrared safe quantity $F$, i.e.\ a 
distribution to be binned in a histogram $\done F$.  For the $m$-parton integral no 
special treatment is mandatory: for a given momentum configuration, $\done F$ can directly 
be evaluated and filled into the corresponding bin.  For the real correction and the 
dipole subtraction functions in the $(m+1)$-parton integral, $F$ has to be evaluated for 
each contribution separately, similar to the phase-space cut.  Again, the problem of 
``missed binnings'' appears, if contributions to a single event end up in more than one bin.

\section{Checks of the implementation}
\label{tests}
In this section a number of tests of the correct implementation of
the subtraction algorithm and of the integration routines are described.
These tests are mainly technical in nature, results relating to truly
physical observables are discussed in the next section, Sec.\ 
\ref{applications}.

\subsection{Explicit comparisons}
Before moving on to technical checks, it is worth stating that a number 
of direct comparisons of individual terms from the program presented here
with those obtained from M.\ Seymour's Fortran code \disent have been 
performed.  The latter is a dedicated program to compute NLO cross sections
for the deep inelastic scattering processes $e^- p\to e^- +jet$, 
$e^- p\to e^- +2jets$ and for electron-positron annihilation to two and 
three jets.  This direct comparison is possible, since \disent uses 
exactly the same subtraction formalism, allowing to compare individual terms
at given phase space points.  All terms listed in the following showed 
full agreement of the two codes, up to the numerical precision.

The comparison included:
\begin{itemize}
\item Dipole subtraction terms for the real correction:\\ 
        all flavour configurations for dipoles with final state emitters 
        and spectators as well as for dipoles with initial state emitters / 
        final state spectators and final state emitters / initial state 
        spectators have been checked.
\item Terms from the finite part of the insertion operator ${\bf I}$,\\ 
        cf.\ Eqs.\ (\ref{Eq:BornTimesI}) and (\ref{Eq:Insertionoperator}).
\item Terms from the insertion operators ${\bf K}$ and ${\bf P}$ for 
        the case of one initial state parton,\\ 
        cf.\ Eq.\ (\ref{fullfiniteexcl}) and the implemented version Eq.\ 
        (\ref{fullfinitreorderd}).
\end{itemize}

Furthermore integrated results of the virtual and real parts of the 
NLO corrections in this subtraction scheme where compared and agreed 
within statistical errors for all accessible processes.

\subsection{Test of convergence for the real ME}
\begin{figure}
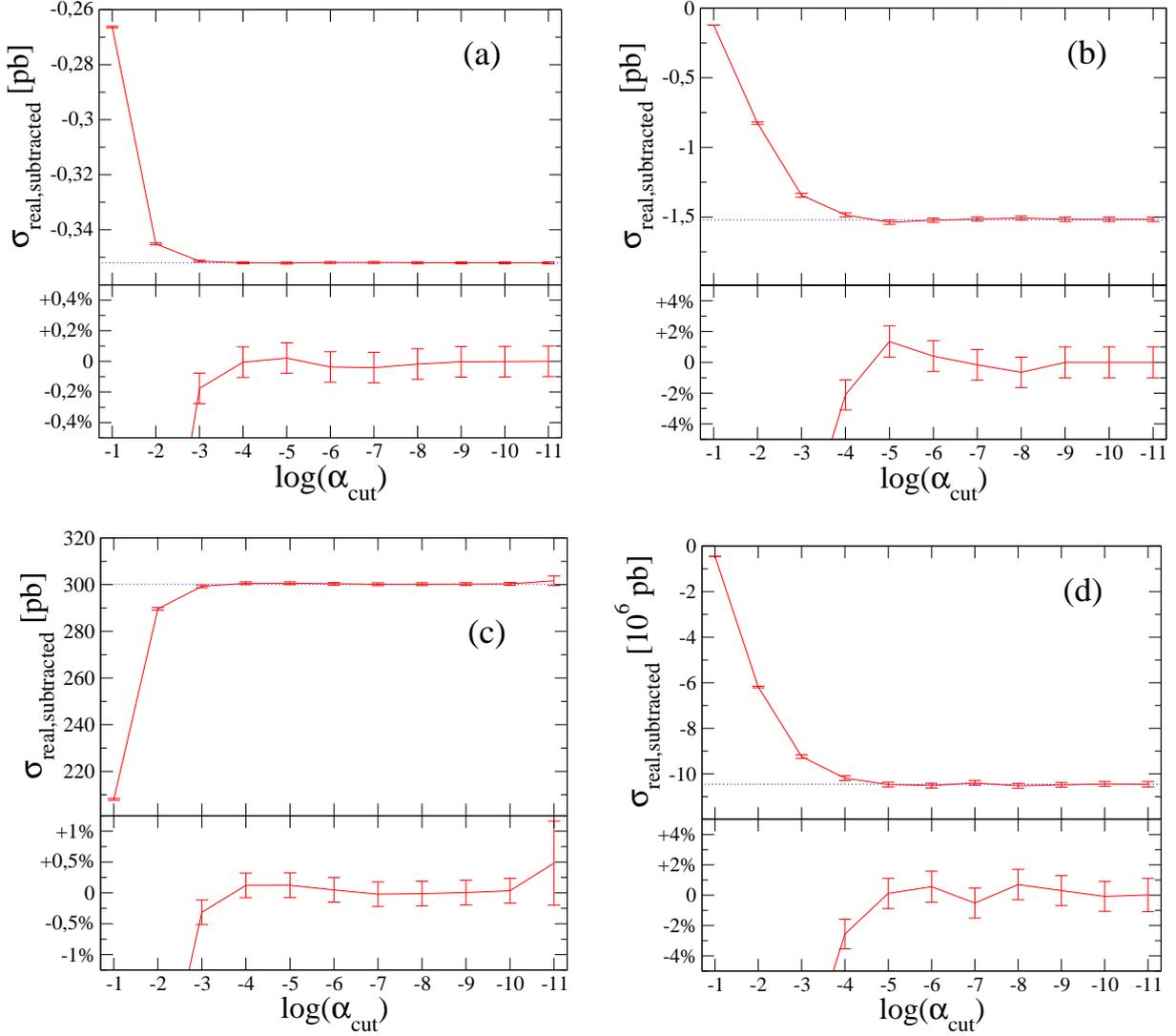

  \begin{center}
    \begin{picture}(480,420)
      \put(5,0){\includegraphics[width=7.6cm]{eje2jXS.eps}}
      \put(238,0){\includegraphics[width=7.6cm]{pp3jXS.eps}}
      \put(0,210){\includegraphics[width=7.7cm]{ee3jXS.eps}}
      \put(240,210){\includegraphics[width=7.6cm]{ee4jXS.eps}}
    \end{picture}
    \caption{Dependence of the subtracted real emission cross section 
        on $\alpha_{\rm cut}$ for
        (a):~$e^-e^+\to 2 jets$; 
        (b):~$e^-e^+\to 3 jets$, both at a CM energy of 100 GeV;
        (c):~$e^-p\to e^-+jet$ with a 50 GeV electron beam and  
             protons at 500 GeV;
        (d):~$pp\to 2 jets$ at a CM energy of 14 TeV. 
        To obtain a well-defined LO cross section for (b) at least two 
        jets with a $k_\perp^{\rm Dur.}>10$ GeV, for (c) a transverse
        energy of the scattered $e^-$ $>10$ GeV and for (d) at least two
        jets with $p_\perp>40$ GeV are required.}
        \label{test_alphacxs} 
  \end{center}
\end{figure}
\begin{figure}
  \begin{center}
    \includegraphics[width=155mm]{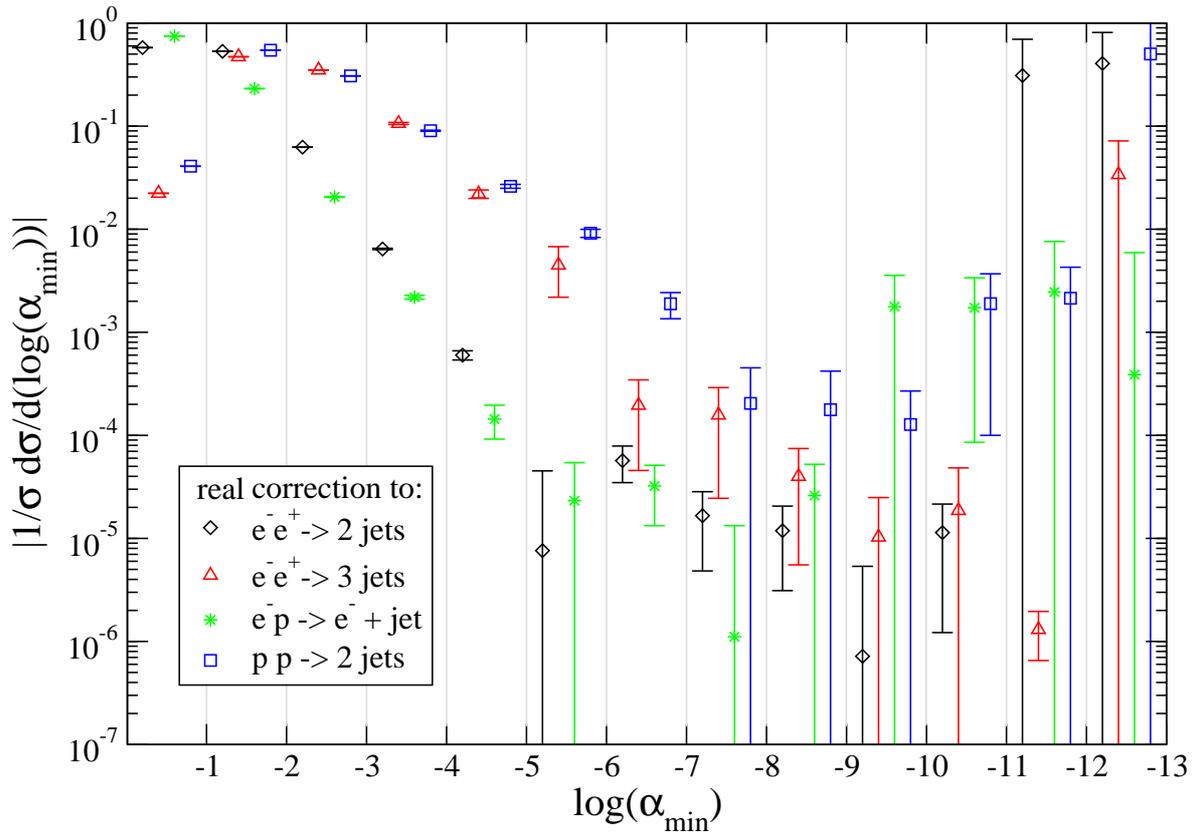}
    \caption{Normalized absolute values of cross sections in bins
        of $\alpha_{\rm min}$. Setups and phase space cuts are the same 
        as in Fig.\ \ref{test_alphacxs}.}
        \label{alphacbin}
  \end{center}
\end{figure}

\label{conv_chk}
An obvious first technical check of the overall package consists of 
testing the convergence behaviour of the dipole subtraction terms close 
to the singular region.  To this end, the $m+1$-parton phase space of 
the regularized real correction part is numerically integrated over.  
The crucial issue is to ensure that the integrand remains finite over 
the full phase space, in addition the performance of the integration 
algorithms deserve some consideration.  

Clearly, for the numerical calculation a small phase space region around 
each singular configuration has to be cut out.  Although the dipole terms 
are expected to become equal to the matrix elements there, technically 
speaking infinite or very large numbers must be subtracted in this region, 
leading to large fluctuations and hence to errors due to the limited 
numerical precision at which the calculation is performed.  Therefore 
a variable $\alpha_{\rm min}$ is introduced, which on the basis of 
kinematic variables of corresponding dipole functions, reads as follows:
\bea
\alpha_{\rm min}=\min_{\rm dipoles} (a_{\rm dipole})\;,
\eea
where
\bea
a_{\rm dipole}&=&\left\{
\begin{matrix}
y_{ij,k}\;\;&{\rm for}\;{\cal D}_{ij,k}\;
{\rm - dipoles\;(final\; state\; emitter,\;final\; state\; spectator)}\\
1-x_{ij,a}\;\;&{\rm for}\;{\cal D}_{ij}^a\;
{\rm - dipoles\;(final\; state\; emitter,\;initial\; state\; spectator)}\\
u_{i}\;\;&{\rm for}\;{\cal D}^{ai}_k\;
{\rm - dipoles\;(initial\; state\; emitter,\;final\; state\; spectator)}\\
\tilde{v}_{i}\;\;&{\rm for}\;{\cal D}^{ai,b}\;
{\rm - dipoles\;(initial\; state\; emitter,\;initial\; state\; spectator)}
\end{matrix}\right.\;.
\eea
This parameter $\alpha$ serves as a cut-off in such a way that for an
externally given parameter $\alpha_{\rm cut}$ kinematic configurations
with $\alpha_{\rm min}<\alpha_{\rm cut}$ are omitted. 

In Fig.\ \ref{test_alphacxs} the dependence of the subtracted cross section 
on $\alpha_{\rm cut}$ for four sets of real correction processes, namely
$e^-e^+\to 3 jets$, $e^-e^+\to 4 jets$, $e^-p\to e^-+2jets$ and 
$pp\to 3 jets$.  All types of dipoles and splitting functions contribute to 
the dipole terms which are necessary to regularize those processes.  It is 
apparent that for $\alpha_{\rm cut}\sim 10^{-5}$ the cross section
stabilizes close to its final value. 

To study the numerical behaviour near the singularity in more detail, in 
Fig.\ \ref{alphacbin} the absolute value of the subtracted cross section, 
binned in intervals of $\alpha_{\rm min}$ is depicted.  For all studied 
processes the contribution to the cross section drops down by at least 
four orders of magnitude with decreasing $\alpha_{\rm min}$ and confirms 
the observations for full subtracted cross sections made before.

The strong increase accompanied with statistical errors of 100\% or larger
for $\alpha_{\rm min}$ values below $10^{-9}-10^{-11}$ signals defects
due to the limited numerical precision ({\it double} precision $\sim10^{-12}$).
One reason is the already mentioned numerical problem when subtracting 
extreme large and almost equal numbers.  Another reason is the precision
of the momentum four-vectors itself, because the precision of the external
particles residing on their $m=0$ mass shell is also limited by the 
numerical precision.  This of course may consequently lead to errors of 
that order in the matrix element calculation.  Thus, Fig.\ \ref{alphacbin} 
allows to determine best choices for $\alpha_{\rm cut}$, somewhere
between $10^{-9}$ and $10^{-11}$.

\subsection{Consistency checks with free parameters}
In section \ref{dtfreedom} ways of modifying the subtraction terms without 
changing the singular behaviour have been discussed.  Such modifications 
can be employed for non-trivial tests of the implementation, since the
modifications will affect both, the real part and the virtual part of the 
NLO cross sections, with their sum remaining constant.  
\begin{figure}
  \begin{center}
    \begin{picture}(480,420)
      \put(0,0){\includegraphics[width=7.55cm]{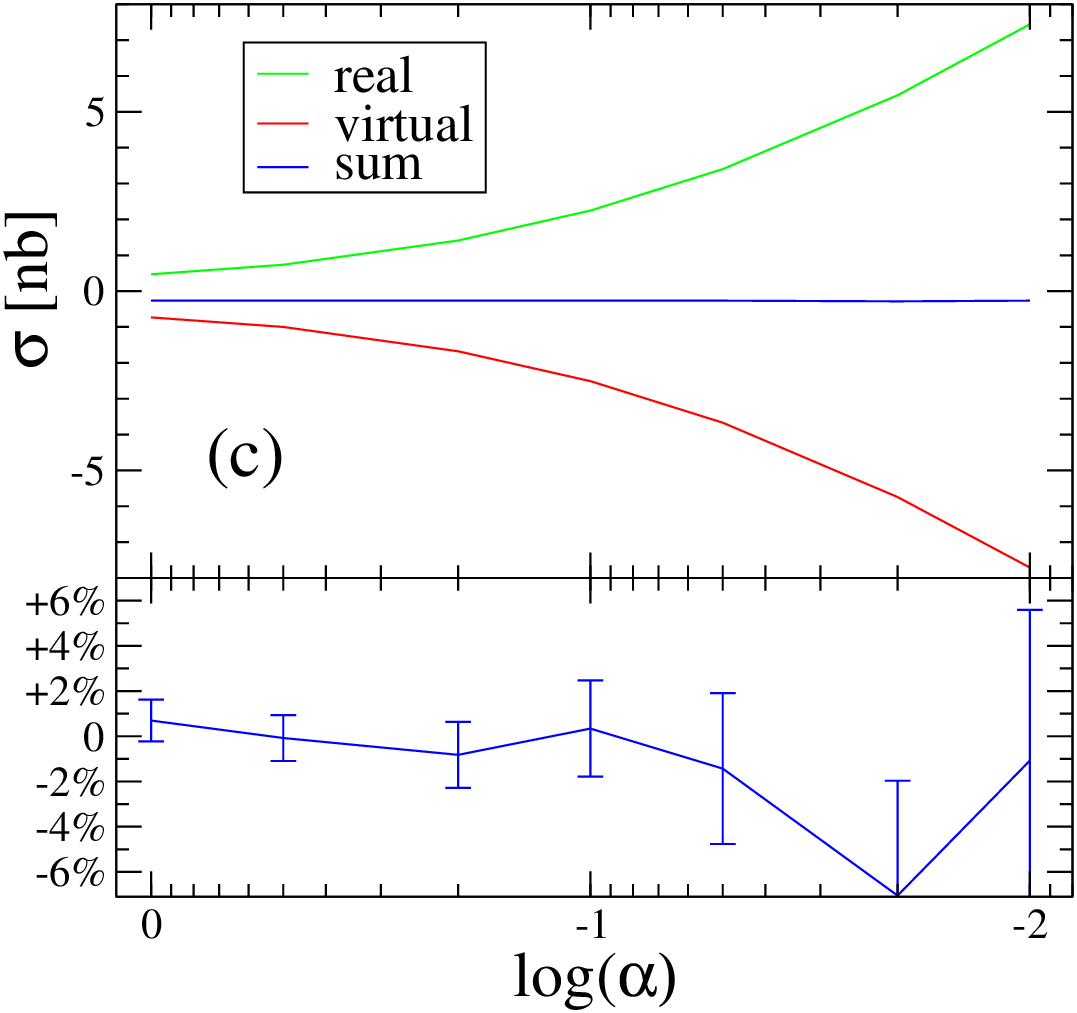}}
      \put(240,0){\includegraphics[width=7.73cm]{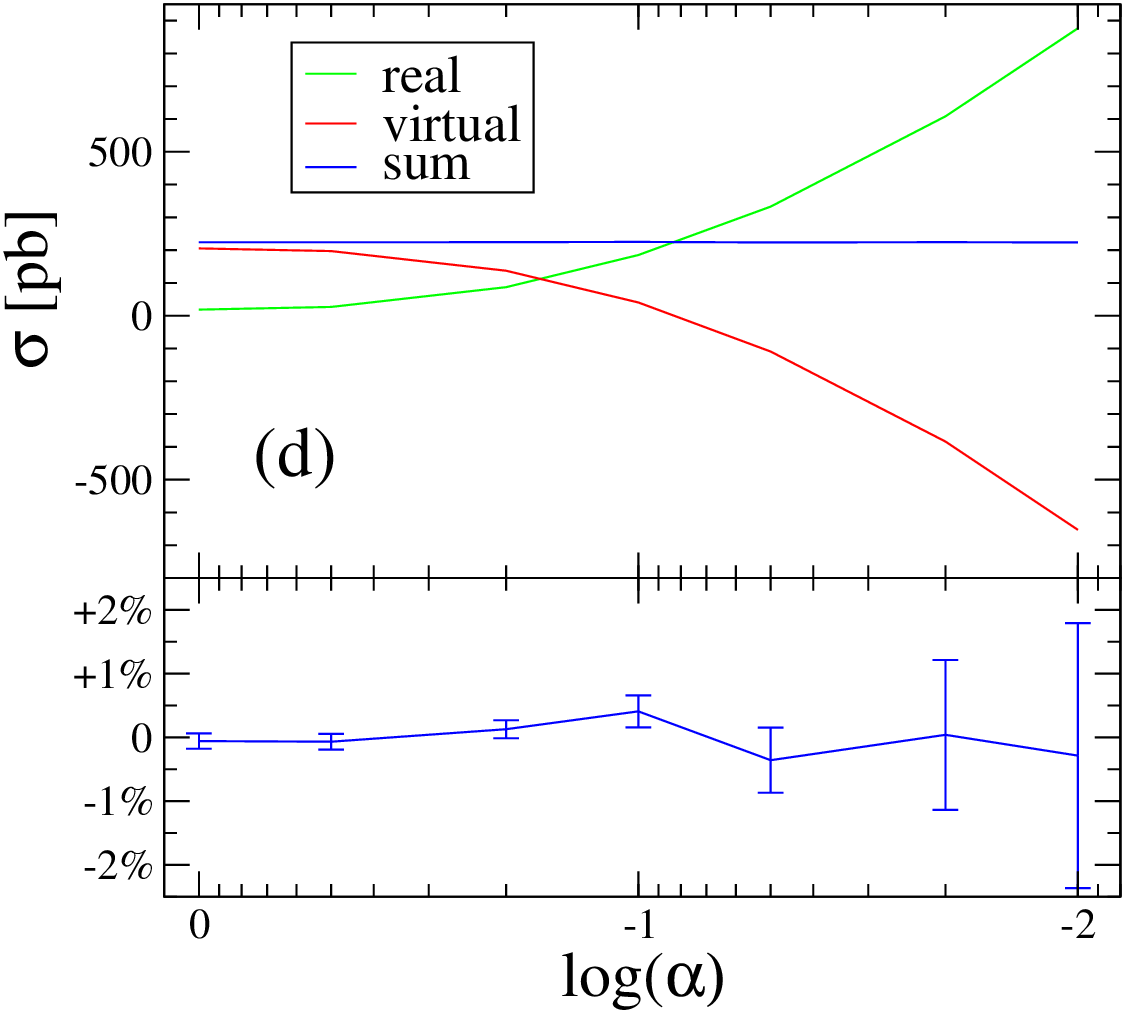}}
      \put(0,217){\includegraphics[width=7.6cm]{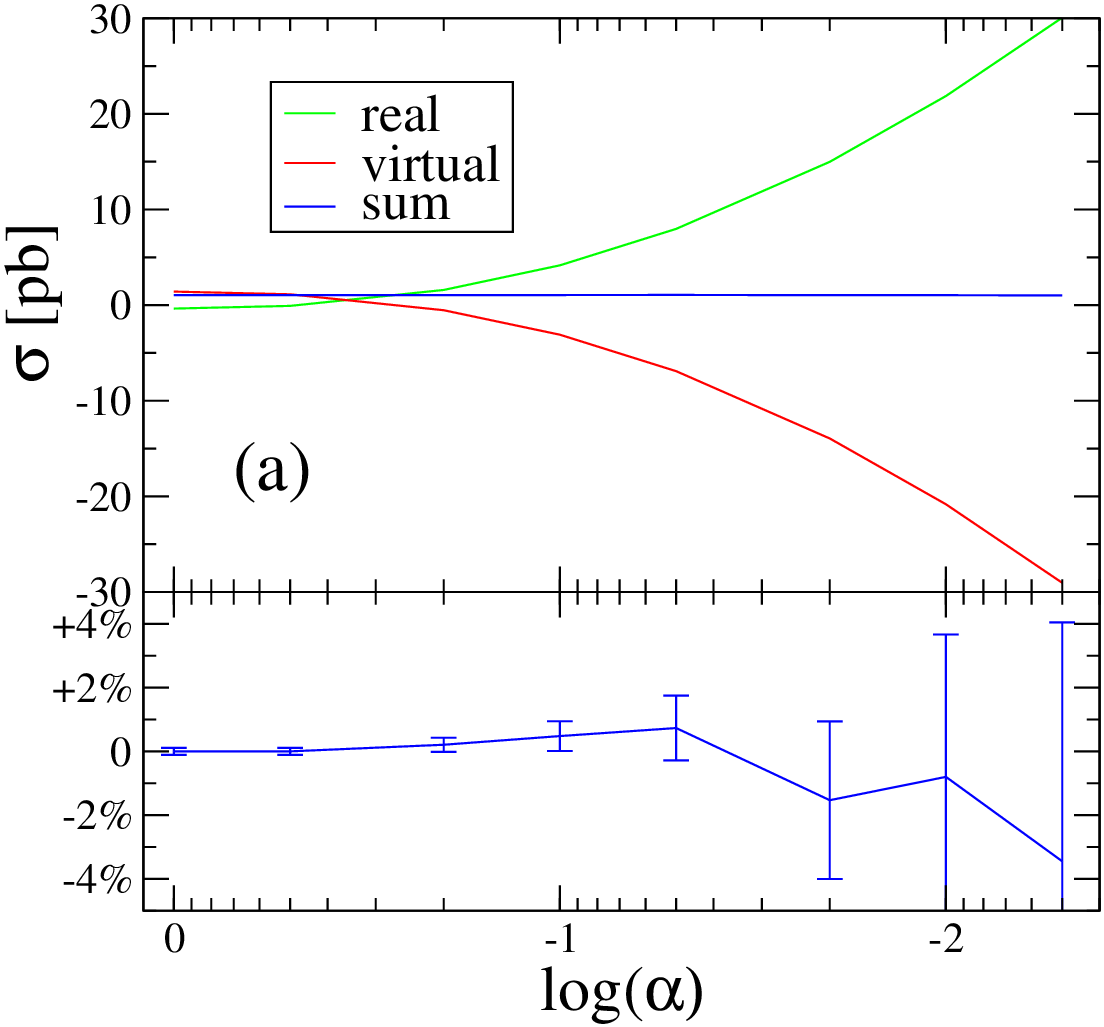}}
      \put(240,217){\includegraphics[width=7.7cm]{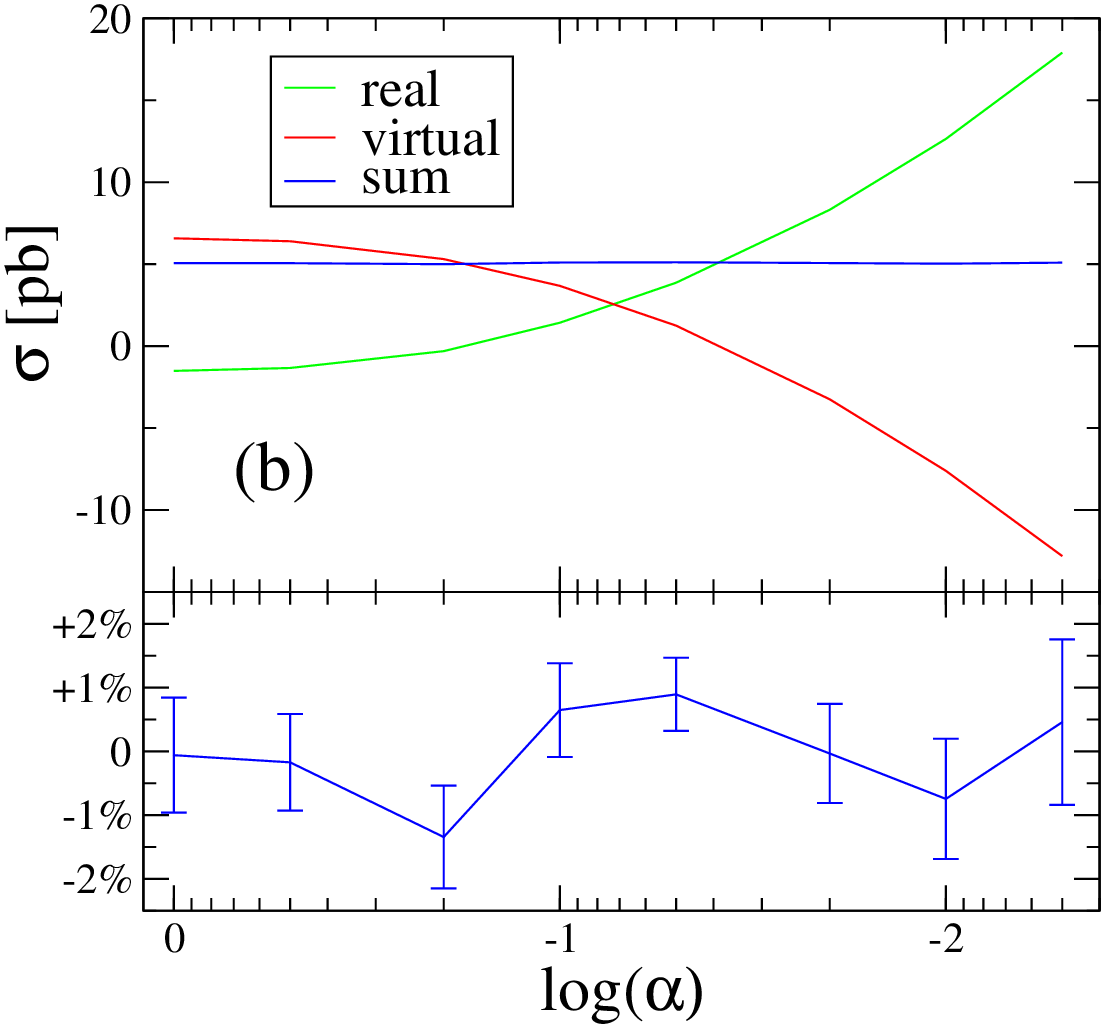}}
    \end{picture}
    \caption{NLO corrections as a function of the parameter $\alpha$ in 
        the definition of the subtraction terms (see Eq.\ (\ref{nagyalpha})) 
        for the total cross sections of (a) $e^-e^+\to 2 jets$, 
        (b) $e^-e^+\to 3 jets$, (c) $e^-p\to e^-+jet$ and (d)
        $p\bar p \to W^- \to e^- \bar\nu_e$.  The results and error bars 
        in the difference plots are determined after calculating 500000 
        phase space points of the real contribution, which typically 
        dominates the total statistical error.}
        \label{test_nlo_alpha} 
  \end{center}
\end{figure}

In Fig.\ \ref{test_nlo_alpha} the total NLO correction for the cross 
sections of $e^-e^+\to 2 jets$, $e^-e^+\to 3 jets$, $e^-p\to e^-+jet$ 
and $p\bar p \to W^- \to e^- \bar\nu_e$ and their real and virtual 
contributions are displayed as functions of the parameter $\alpha$, as 
introduced in section \ref{dtfreedom}.  The fact that the sum remains 
constant within statistical errors provides a non-trivial confirmation of
the correct implementation of the algorithm.  It should be noted here
that the calculation of the cross section of the processes under 
consideration invokes all types of dipole functions as well as the most 
general case of the insertion operators from the integrated dipole terms.

By using the same number of phase space points for each integral and 
comparing statistical errors, it can be seen that this parameter can also
be used to optimize the numerical behaviour.  Clearly, best results are
obtained if the values of the virtual and  real contributions are both as 
small as possible, thus reducing the size of the fluctuations. 
It should be noted that the error bars in Fig.\ \ref{test_nlo_alpha}
are given not including the leading order part of the cross section. 
Since relative errors for the
latter can be expected to be much smaller if evaluated for the same number
of phase space points, the (relative) statistical error for the full NLO cross section
will be significantly reduced.

\section{First physical applications}
\label{applications}
In this section, some simple applications will demonstrate the performance 
of the dipole subtraction procedure, as implemented, for the calculation
of physically relevant observables.  The born matrix elements, dipole 
subtraction terms to regularize the real correction and corresponding finite 
terms to be added to the virtual correction were generated automatically 
by \amegic.  The one-loop amplitudes have been explicitly implemented 
for the considered processes. 

\subsection{Three-jet observables at LEP}
\begin{figure}
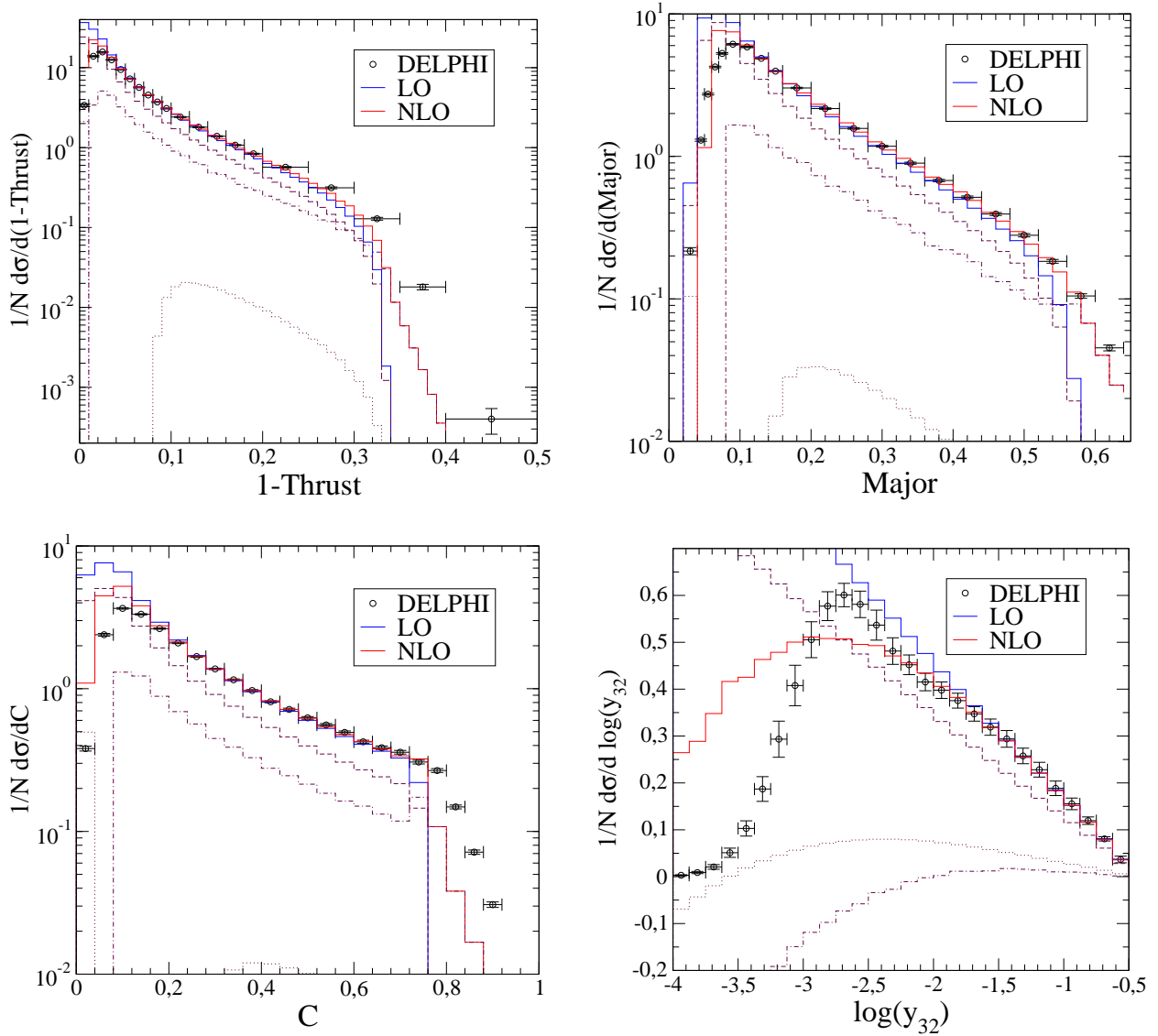

  \begin{center}
    \begin{picture}(480,420)
      \put(0,3){\includegraphics[width=7.7cm]{C_Parameter_normalized.eps}}
      \put(240,0){\includegraphics[width=7.9cm]{y32_normalized.eps}}
      \put(0,222){\includegraphics[width=7.8cm]{thrust_normalized.eps}}
      \put(240,220){\includegraphics[width=7.7cm]{major_normalized.eps}}
    \end{picture}
    \caption{The event shape observables 1-Thrust, Major, C-parameter
        and the Durham $3\to2$ jet rate at LEP I compared to measurements 
        by DELPHI \cite{Abreu:1996na}.  The LO and the NLO predictions 
        have been normalized to the data separately in a region, where 
        agreement can be expected.  The dashed, dashed-dotted and the 
        dotted lines are the Born, real and virtual contribution to the 
        NLO cross sections, respectively.}
        \label{LEP_nlo} 
  \end{center}
\end{figure}
To compute three jet cross section at next-to-leading order the one-loop 
matrix element given in \cite{Ellis:1980wv} has been implemented.  The 
expression given there is averaged over the direction of incoming momenta,
which is sufficient for observables that are not correlated to the beam 
direction.

In Fig.\ \ref{LEP_nlo} LO and NLO predictions are displayed for observables 
sensitive to ${\cal O}(\alpha_S)$.  In particular, the event shape 
observables 1-Thrust, Major, C-parameter and the Durham $3\to2$ jet rate 
are compared with measurements performed at LEP on the $Z^0$-peak by 
DELPHI \cite{Abreu:1996na}.  All data are normalized to unity.  The 
normalisation for the calculated cross section, however, is somewhat 
complicated.  This is because in the calculation three-jet events are 
required in each case, translating into the necessity to apply a phase 
space cut.  On the other hand, the data are more inclusive and also include
comparably soft regions, where fixed-order perturbation theory is known to 
fail and must be supported by resummation techniques.  The normalisation for
the calculations has thus be chosen such that it agrees with data in 
the ``safe'' regions.  This exposes the differences between LO and NLO
calculations best.  As a consequence, the corresponding normalisation 
factor of both calculations is not identical.  From the results of Fig.\ 
\ref{LEP_nlo} it can be deduced that for all observables the range 
described sufficiently well by the calculation is extended for the NLO 
calculations.  For both, the region described by soft physics (left side 
in all plots) as well as phase space regions populated by additional hard 
QCD radiation (right side in all event shape plots) the prediction has been
improved. 

\subsection{DIS: $e^-p\to e^-+jet$}
\begin{figure}
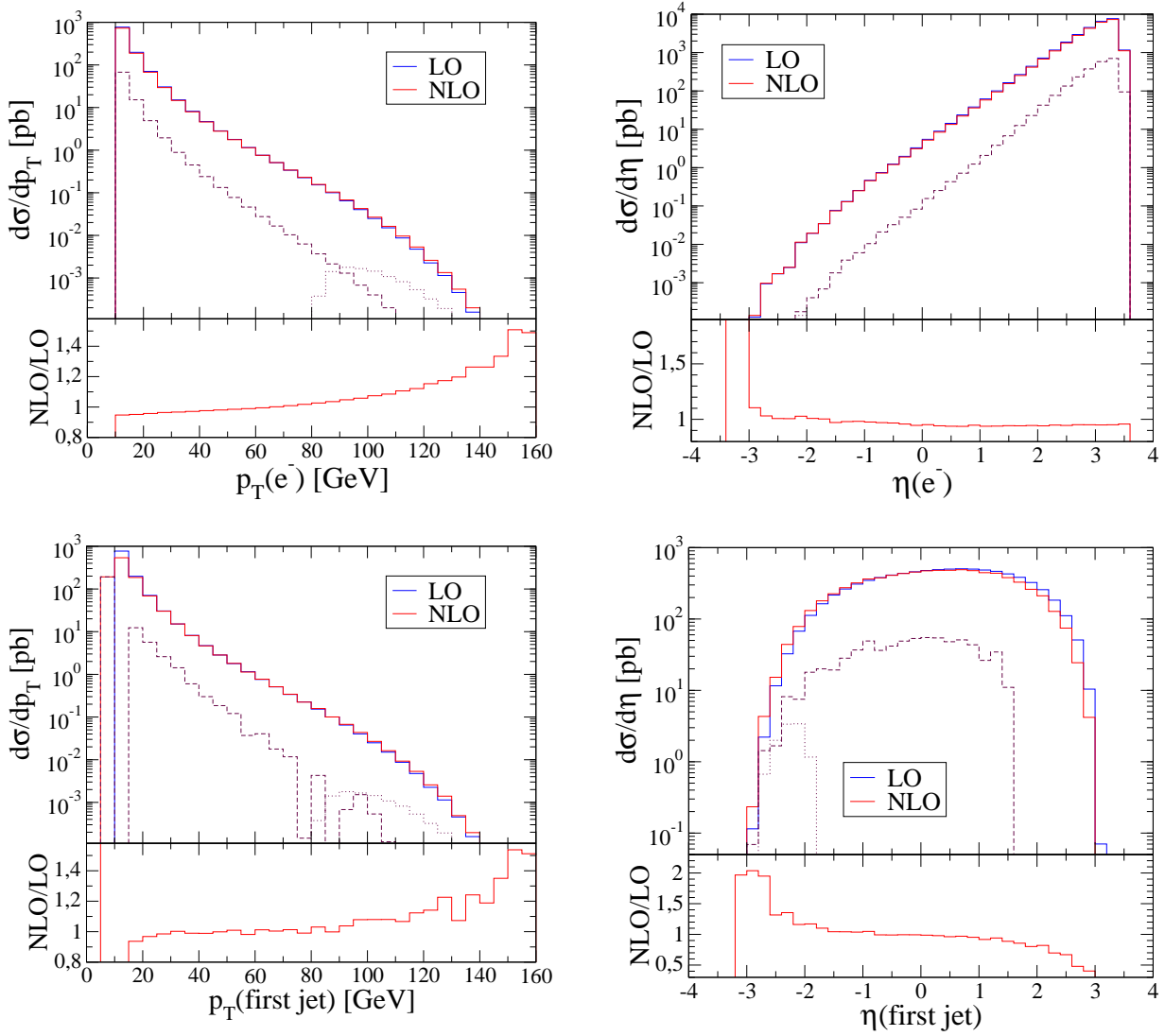

  \begin{center}
    \begin{picture}(480,420)
      \put(0,-8){\includegraphics[width=7.6cm]{PTjet_Z.eps}}
      \put(240,0){\includegraphics[width=7.6cm]{eta_jet.eps}}
      \put(0,212){\includegraphics[width=7.6cm]{PTe_Z.eps}}
      \put(240,212){\includegraphics[width=7.6cm]{eta_e.eps}}
    \end{picture}
    \caption{Distribution of transverse momentum (left plots) and rapidity, 
        defined in the beam CM frame (right plots), of the scattered 
        electron and of the hardest jet in deep inelastic scattering, 
        calculated at leading order and next-to-leading order.  The
        CM-energy has been taken as $\sqrt{10^5}$ GeV.  A phase space cut 
        on the electron ($p_T>10 GeV$) has been applied.  For the rapidity 
        distribution of the first jet a $p_T>15 GeV$ has been required.
        Dashed and dotted lines denote the real and the virtual corrections 
        to the Born cross section, respectively.  The lower panels of each 
        plot show the ratio between the leading order and the next-to-leading 
        order results.}
        \label{DIS_nlo} 
  \end{center}

\end{figure}
The one loop matrix element for this process is given by the well known
expression
\bea
|M|^2_{(1-loop)}&=&
|M|^2_{(born)}\frac{C_F\alpha_S}{2\pi}\frac1{\Gamma(1-\epsilon)}
\left(\frac{4\pi\mu^2}{Q^2}\right)^\epsilon
\left\{-\frac2{\epsilon^2}-\frac3\epsilon-8+{\cal O}(\epsilon)\right\}\;,
\label{eqeq_loop}
\eea
where $Q^2=-q^2>0$ with $q$ the momentum transfer between the electron 
and the proton. 
 
Fig.\ \ref{DIS_nlo} shows differential cross sections w.r.t.\ the 
transverse momentum and rapidity of the scattered electron and the 
hardest jet at leading and at next-to-leading order.  The CM-energy 
has been taken as $\sqrt{10^5}$ GeV, corresponding to a 50 GeV electron 
beam and a 500 GeV proton beam.  The parton distribution function 
CTEQ6M \cite{Pumplin:2002vw} has been employed, factorisation and 
renormalisation 
scales have been fixed to $Q^2$.  A phase space cut on the electron of 
$p_T>10 GeV$ has been imposed.  The NLO correction for this setup is 
comparably small, for the total cross section it is of the order of $5\%$ 
and negative.  The ratios of NLO and LO calculation, however, are not 
constant for all observables.  At NLO the cross section rises for  
increasing momentum transfer, up to a correction of 40\% for transverse
momenta of electron and jet of the order of 150 GeV.

\subsection{$W^-$ production at Tevatron}
\begin{figure}
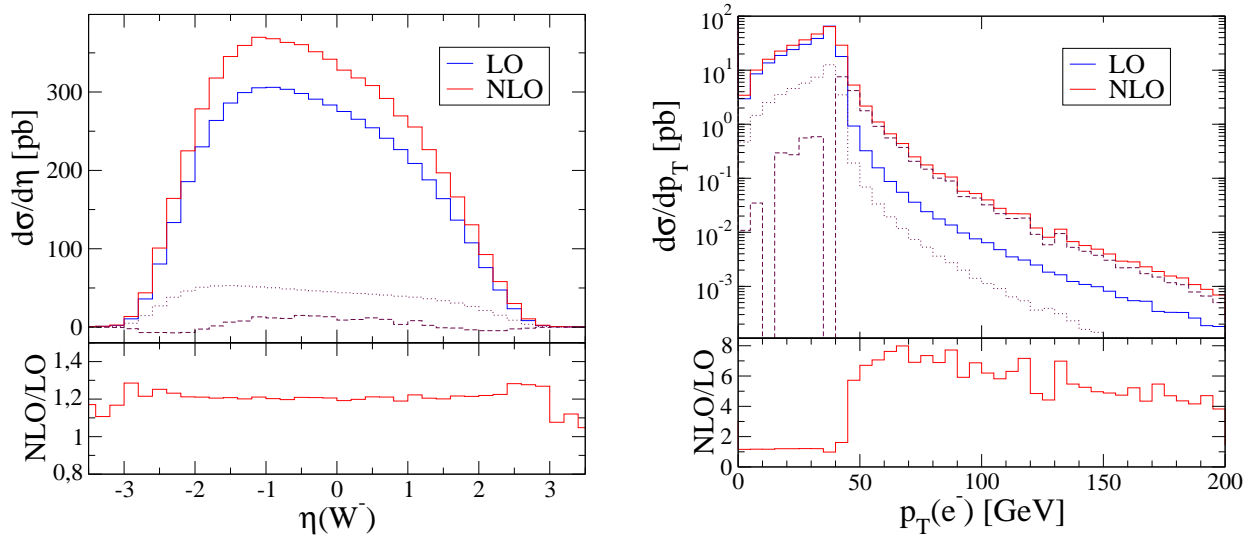

  \begin{center}
    \begin{picture}(480,200)
      \put(0,0){\includegraphics[width=7.6cm]{DY_yW.eps}}
      \put(240,0){\includegraphics[width=7.8cm]{DY_PTe.eps}}
    \end{picture}
    \caption{Rapidity distribution of the $W^-$-boson (left plot) 
        and transverse momentum of the electron for the process
        $p\bar p\to W^- \to e^- \bar\nu_e$ at Tevatron Run II,
        calculated at leading order and next-to-leading order.
        Dashed and dotted lines denote the real and the virtual corrections 
        to the Born cross section, respectively.  The lower panels of each 
        plot show the ratio between the leading order and the next-to-leading 
        order results.}
        \label{DY_nlo} 
  \end{center}
\end{figure}
The one-loop virtual contribution to this process can be obtain by crossing 
relations from Eq.~(\ref{eqeq_loop}) and is given by
\bea
|M|^2_{(1-loop)}&=&
|M|^2_{(born)}\frac{C_F\alpha_S}{2\pi}\frac1{\Gamma(1-\epsilon)}
\left(\frac{4\pi\mu^2}{Q^2}\right)^\epsilon
\left\{-\frac2{\epsilon^2}-\frac3\epsilon-8+\pi^2+{\cal O}(\epsilon)\right\}\;,
\eea
where now $Q^2=\hat s$, the CM energy squared of the incoming partons.

Fig.\ \ref{DY_nlo} shows cross sections for Tevatron Run II,
differential in the rapidity of the $W^-$-boson and the transverse 
momentum of the electron, respectively.
The parton distribution function CTEQ6M \cite{Pumplin:2002vw} has been employed, 
factorisation and renormalisation scales have been fixed to $m_W^2$.

The total and differential cross sections are in full agreement
with predictions obtained using the next-to-leading order parton level
generator MCFM \cite{Campbell:2002tg}.
\section{Conclusions and outlook}
In this publication a fully automated implementation of the Catani-Seymour
dipole formalism in the framework of the matrix element generator \amegic
has been presented.  It allows to automatically generate the process-dependent
real correction terms for given Born cross sections with massless external 
particles and the corresponding real subtraction terms.  The integration of 
the subtracted real correction terms is performed automatically with a 
multi-channel method, giving rise to an appreciable convergence.  The
implementation has carefully been checked for correctness, invoking 
consistency checks with free finite terms which may be added to the 
subtraction terms.  Through the explicit inclusion of virtual terms a 
parton-level calculator is so available.  

In the future, the code will be further updated to include massive external
particles and to provide a full parton-level generator at NLO. 

\section{Acknowledgments}
The authors would like to thank M.\ Seymour for fruitful discussions 
and numerous comparisons to the code \disent.  Financial support by
BMBF and the Marie Curie research training network MCnet (contract number
MRTN-CT-2006-035606) is gratefully acknowledged.  Furthermore, T.\ G.\ would
like to thank the Marie Curie Fellowship program for Early Stage Training 
hosted by CERN for financial support over an extended period of 11 months
and the CERN theory group for kind hospitality.

\appendix
\section{Insertion operators etc.}
\label{appendix:littlefuncs}

In this appendix, the ingredients of the master equation Eq.\ (\ref{fullfiniteexcl}),
\bea
\done\sigma^{\tilde{A}}_{ab}(p_a,p_b)+d\sigma^C_{ab}(p_a,p_b,\mu_F^2)
&=&\;\;\;
\left[\done\sigma^B_{ab}(p_a,p_b)\times{\bf I}(\epsilon)\right]\nnb\\
&&
+\sum_{a^\prime}\int_0^1\done x
      \left[\left({\bf K}^{a,a^\prime}(x)+
                  {\bf P}^{a,a^\prime}(xp_a,x;\mu_F^2)\right)
            \times \done\sigma^B_{a^\prime b}(xp_a,p_b)\right]\nnb\\
&&
+\sum_{b^\prime}\int_0^1\done x
      \left[\left({\bf K}^{b,b^\prime}(x)+
                  {\bf P}^{b,b^\prime}(xp_b,x;\mu_F^2)\right)
            \times \done\sigma^B_{a b^\prime}(p_a,xp_b)\right]\;,\nnb\\
\eea
will be repeated.  $a$ and $b$ specify initial state partons, and the sum runs 
over all accessible $a^\prime$ and $b^\prime$ occurring in the PDF.  The insertion 
operator ${\bf I}$ is given by
\bea
{\bf I}(\{p\};\epsilon)
=
-\frac{\alpha_S}{2\pi}\frac1{\Gamma(1-\epsilon)}
 \sum_I\frac1{{\bf T}_I^2}{\cal V}_I(\epsilon)
 \sum_{I\neq J}{\bf T}_I\!\cdot\!{\bf T}_J
       \left(\frac{4\pi\mu^2}{2p_Ip_J}\right)^\epsilon\;,
\eea
cf.\ Eq.\ (\ref{Eq:Insertionoperator}), and again the indices $I$ and $J$ run over all 
initial and final state partons, while the universal functions ${\cal V}_I(\epsilon)$, 
encoding the singularity structure, merely depend on the flavour of $I$ and read
\bea\label{Eq:UniversalV'}
{\cal V}_I(\epsilon)
&=&
{\bf T}_I^2\left(\frac1{\epsilon^2}-\frac{\pi^2}{3}\right) +
\gamma_I\left(\frac{1}{\epsilon}+1\right) + K_I + {\cal O}(\epsilon)\,,
\eea
cf.\ Eq.\ (\ref{Eq:UniversalV}).  The individual $\gamma_I$ and $K_I$ will
be listed in Eqs. (\ref{Eq:gammas}) and (\ref{Eq:KIs}).  

The factorisation scale dependent terms are proportional to insertion operators
${\bf P}^{a,a^\prime}(\{p\},xp_a,x;\mu_F^2)$, which read
\bea\label{Eq:BFPs}
{\bf P}^{a,a^\prime}(\{p\},xp_a,x;\mu_F^2)
&=&
\frac{\alpha_S}{2\pi}P^{aa^\prime}(x)\frac{1}{{\bf T}_{a^\prime}^2}\,
\sum\limits_{I\ne a^\prime}{\bf T}_I\cdot {\bf T}_{a^\prime}\ln\frac{\mu_F^2}{2xp_a\cdot p_I}\,.
\eea
The regularized Altarelli-Parisi kernels $P^{ab}(x)$ are listed in 
Eq.\ (\ref{Eq:APKernels}).

The factorisation-scheme dependent terms are proportional to the initial-state insertion
operators ${\bf K}$.  For one initial-state hadron only, this operator reads
\bea\label{Eq:K4OneIS}  
{\bf K}^{a,a^\prime}(x) = \frac{\alpha_S}{2\pi}
\left\{\bar K^{aa^\prime}(x)-K^{aa^\prime}_{F.S}(x)+
\delta^{aa^\prime}\sum\limits_i\frac{\gamma_i {\bf T}_i\cdot{\bf T}_a^\prime}{{\bf T}_i^2}
\left[\left(\frac{1}{1-x}\right)_++\delta(1-x)\right]
\right\}\,,
\eea
with the functions $K^{aa^\prime}_{\rm F.S.}(x)$ and $\bar K^{aa^\prime}(x)$ given below, 
cf.\ Eqs.\ (\ref{Eq:BarKofX}) and (\ref{Eq:FSKofX}), and with the $\gamma_i$ listed in Eq.\ 
(\ref{Eq:gammas}).  Note that the subscript ``F.S.'' denotes the factorisation scheme.  
For two initial state partons, the initial-state insertion operator is given by Eq.\ 
(\ref{Eq:ISinsertion}),
\bea\label{Eq:ISinsertion'}
{\bf K}^{a,a^\prime}(x)
&=&
\frac{\alpha_S}{2\pi}\left\{\bar{K}^{aa^\prime}(x)-K^{aa^\prime}_{\rm F.S.}(x)\right.\nnb\\
&&\hspace*{6mm}\left.
+\delta^{aa^\prime}\sum_i{\bf T}_i\!\cdot\!{\bf T}_a\frac{\gamma_i}{{\bf T}_i^2}
\left[\left(\frac1{1-x}\right)_++\delta(1-x)\right]-
\frac{{\bf T}_{b}\!\cdot\!{\bf T}_{a^\prime}}{{\bf T}_a^2}\tilde{K}^{a,a^\prime}(x)
\right\}\;,
\eea
with the functions $\tilde K^{aa^\prime}(x)$ given in Eq.\ (\ref{Eq:TildeKofX}).

The $\gamma_I$ and $K_I$ occurring in Eqs. (\ref{Eq:UniversalV'}) are related to 
integrals of the Altarelli-Parisi kernels listed below, Eq.\ (\ref{Eq:APKernels}), and read
\bea\label{Eq:gammas}
\gamma_q = \gamma_{\bar q} = \frac32 C_F\;,\;\;\;
\gamma_g = \frac{11}{6} C_A - \frac23 T_RN_f
\eea
and
\bea\label{Eq:KIs}
K_q = K_{\bar q} = \left(\frac72-\frac{\pi^2}{6}\right)\,C_F\;,\;\;\;
K_g = \left(\frac{67}{18}-\frac{\pi^2}{6}\right)\,C_A-\frac{10}{9}T_RN_f\;,
\eea
respectively.  The Altarelli-Parisi kernels emerging in the factorisation-scale dependent 
terms of Eq.\ (\ref{Eq:BFPs}) are
\bea\label{Eq:APKernels}
P^{qg}(x) = P^{\bar qg}(x) &=& C_F\frac{1+(1-x)^2}{x}\nnb\\ 
P^{gq}(x) = P^{g\bar q}(x) &=& T_R\left[x^2+(1-x)^2\right]\nnb\\
P^{qq}(x) = P^{\bar q\bar q}(x)
                           &=& C_F\left(\frac{1+x^2}{1-x}\right)_+\nnb\\
P^{gg}(x) &=& 2C_A\left[\left(\frac{1}{1-x}\right)_++\frac{1-x}{x}
                        -1+x(1-x)\right]
                +\delta(1-x)\left[\frac{11}{6} C_A - \frac23 T_RN_f\right]\,.\nnb\\
\eea
The functions $\bar K^{ab}(x)$ are explicitly given as
\bea\label{Eq:BarKofX}
\bar K^{q\bar q}(x) = \bar K^{\bar qq}(x) &=& 0\;,\nnb\\ 
\bar K^{qg}(x) = \bar K^{\bar qg}(x) &=& 
        P^{qg}(x)\ln\frac{1-x}{x} + C_Fx\;,\nnb\\
\bar K^{gq}(x) = \bar K^{g\bar q}(x) &=& 
        P^{gq}(x)\ln\frac{1-x}{x} + 2T_Rx(1-x)\;,\nnb\\
\bar K^{qq}(x) = \bar K^{\bar q\bar q}(x) &=& 
        \;\;\;C_F\left[\left(\frac{2}{1-x}\ln\frac{1-x}{x}\right)_+ -
                 (1+x)\ln\frac{1-x}{x}+(1-x)\right]\nnb\\
                                        &&
        -\delta(1-x)(5-\pi^2)C_F\;,\nnb\\
\bar K^{gg}(x) &=& 
        \;\;\;2C_A\left[\left(\frac{1}{1-x}\ln\frac{1-x}{x}\right)_+ +
                 \ln\frac{1-x}{x}\left(\frac{1-x}{x}-1+x(1-x)\right)\right]\nnb\\
                                        &&
        -\delta(1-x)\left[\left(\frac{50}{9}-\pi^2\right)C_A
                 -\frac{16}{9}T_RN_f\right]\;,
\eea
whereas the functions $\tilde K^{ab}(x)$ read
\bea\label{Eq:TildeKofX}
\tilde K^{q\bar q}(x) = \tilde K^{\bar qq}(x) &=& 0\;,\nnb\\ 
\tilde K^{qg}(x) = \tilde K^{\bar qg}(x) &=& 
        P^{qg}(x)\ln(1-x)\;,\nnb\\
\tilde K^{gq}(x) = \tilde K^{g\bar q}(x) &=& 
        P^{gq}(x)\ln(1-x)\;,\nnb\\
\tilde K^{qq}(x) = \tilde K^{\bar q\bar q}(x) &=& 
        C_F\left[\left(\frac{2}{1-x}\ln(1-x)\right)_+-\frac{\pi^2}{3}\delta(1-x)
                -(1+x)\ln(1-x)\right]\;,\nnb\\
\tilde K^{gg}(x) &=& 
        \;\;\;C_A\left[\left(\frac{2}{1-x}\ln(1-x)\right)_+-\frac{\pi^2}{3}\delta(1-x)\right.
\nnb\\
&&\left.
        +2\left(\frac{1-x}{x}-1+x(1-x)\right)\ln(1-x)\right]\;.
\eea
Finally, the factorisation-scheme dependent terms are given through
\bea\label{Eq:FSKofX}
K^{ab}_{\bar MS}(x) &=& 0\;,\nnb\\
K^{qq}_{\rm DIS} = K^{\bar q\bar q}_{\rm DIS} &=& 
        C_F\left[\frac{1+x^2}{1-x}\left(\ln\frac{1-x}{x}-\frac34\right)
                 +\frac{9+5x}{4}\right]_+\;,\nnb\\
K^{gq}_{\rm DIS} = K^{g\bar q}_{\rm DIS} &=& 
        T_R\left[\left(x^2+(1-x)^2\right)\ln\frac{1-x}{x}
                 +8x(1-x)-1\right]\;,\nnb\\
K^{qg}_{\rm DIS} = K^{\bar qg}_{\rm DIS} &=& -K^{qq}_{\rm DIS}\;,\nnb\\
K^{gg}_{\rm DIS} &=& -2N_fK^{gq}_{\rm DIS}\;,\nnb\\
K^{\bar qq}_{\rm DIS} = K^{q\bar q}_{\rm DIS} &=& 0\;.
\eea
\bibliographystyle{amsunsrt_mod}  
\bibliography{bibliography}

\end{document}